\documentclass{article}
\usepackage[utf8]{inputenc}
\usepackage[a4paper, total={6in, 8in}]{geometry}
\usepackage{setspace}

\usepackage{amssymb}
\usepackage{fullpage}
\usepackage{enumerate}
\usepackage{amsthm}
\usepackage{amsmath,amssymb}
\usepackage{mathtools}
\usepackage{graphics,graphicx}
\usepackage{subcaption}
\usepackage{booktabs}
\usepackage{multirow}
\usepackage{multicol}
\usepackage{makecell}
\usepackage{fancyhdr}
\usepackage{textcomp}
\usepackage{setspace}
\usepackage{amsfonts}
\usepackage{upgreek}
\usepackage{dsfont}
\usepackage{natbib}
\usepackage{hyperref}
\usepackage{tikz}
\usepackage{array}
\usepackage{float}
\usepackage{dsfont}
\usepackage{booktabs} % For formal tables
\usepackage{cleveref}
\usepackage{siunitx}
\usepackage{soul}

\usepackage{graphicx}
\usepackage{subcaption}
\usepackage{float}
\usepackage{dcolumn} % For decimal alignment
\usepackage{bm}
\usepackage[ruled]{algorithm2e} % For algorithm

\SetAlFnt{\small}
\SetAlCapFnt{\small}
\SetAlCapNameFnt{\small}
\SetAlCapHSkip{0pt}
\IncMargin{-\parindent}

\usepackage[T1]{fontenc}

\usepackage{xcolor}
\usepackage{soul}

\newcommand{\sigmamu}{\ensuremath{\sigma_\mu}}

\newcommand{\mui}{\ensuremath{\mu}}

\newtheorem{definition}{Definition}

\newtheorem{proposition}{Proposition}
\newtheorem{lemma}{Lemma}

\newtheorem{corollary}{Corollary}

\title{Second-degree Price Discrimination: Theoretical Analysis, Experiment Design, and Empirical Estimation}
\author{Soheil Ghili\thanks{Yale University, email address: soheil.ghili@yale.edu}, K. Sudhir\thanks{Yale University, email address: k.sudhir@yale.edu}, Nitish Jain\thanks{London Business School, email address: njain@london.edu}, Ankur Garg\thanks{Air India, email address: ankur1808@gmail.com}}

\date{\today}

\begin{document}
\pagenumbering{gobble} % Suppress page numbering
\maketitle

\onehalfspacing

\begin{abstract}
  We build on theoretical results from the mechanism design literature to analyze empirical models of second-degree price discrimination (2PD). We show that for a random-coefficients discrete choice (``BLP'') model to be suitable for studying 2PD, it must capture the covariance between two key random effects: (i) the “baseline” willingness to pay (affecting all product versions), and (ii) the perceived differentiation between versions. We then develop an experimental design that, among other features, identifies this covariance under common data constraints in 2PD environments. We implement this experiment in the field in collaboration with an international airline. Estimating the theoretically motivated empirical model on the experimental data, we demonstrate its general applicability to 2PD decisions. We also show that test statistics from our design can enable qualitative inference on optimal 2PD policy even before estimating a demand model. Our methodology applies broadly across second-degree price discrimination settings.\end{abstract}

\newpage

\setcounter{page}{1} % Reset page counter to 1
\pagenumbering{arabic} % Start Arabic numbering

%\hl{Some general notes from Sudhir: I have sprinkled some highlights across paper, but happy to edit a separate version for more detailed edits.

%1. May be helpful to have a table of notation--to follow. It may also be useful in propositions to mention the description rather than just notation to make paper easier to read. Intuition will also become stronger.

%2. It may be good to have terms such as partial and total market tests defined in introduction as this is non-standard.

%}

%looks like you have changed the notation to $k=1,2$. I think this is a good idea, and people should be able to easily see that 2 is better than 1 in the labeling.

%May be use some plain english terminology and repeat it consistently when using notation??

%Heterogeneity in baseline value. Different consumers have different overall willingness to pay for product (or service).

%Heterogeneity in upgrade value. Among those who buy, some care a lot more than others about the ``premium'' features (timing, quality, bundle, etc.).

%Covariance of the two. The degree to which high‑value customers also care more (or less) about the premium option is what drives whether two  tier pricing helps at all.

%\newpage

\section{Introduction}
Second-degree price discrimination (“2PD”) is ubiquitous. Its common—and overlapping—forms range from quality-based tiering/versioning of products (e.g., Macbook Pro vs. Macbook Air, or tiers of digital products such as cloud services or software), quantity discounts, varying prices based on purchase timing (e.g., a discount for early booking of accommodation), and product bundling.
The theoretical problem of the optimal selling mechanism for a monopolist in 2PD settings is a case of the ``screening problem''. The literature on screening is vast, especially in the case of product versioning (e.g., different memory levels for smart phones) and bundling (e.g., travel packages). As a number of studies in the screening literature suggest, the monopolist’s optimal mechanism often depends on certain properties of the joint distribution on consumers’ valuations of the different products \citep{anderson2009price,haghpanah2019pure,ghili2023characterization,yang2021costly,yang2023nested}. Hence, to execute the optimal mechanism, a requirement is knowledge of this joint distribution by the designer/seller. Sellers in real-world 2PD settings, however, often lack such information. This can hinder the use of screening theory for 2PD decisions unless it is directly connected to applied settings via an empirical framework. 

This paper’s objective is to make the theory of screening empirically applicable by mechanism designers who do not a priori possess the required information on the joint distribution of the valuations. To this end, we accomplish three objectives. First, we develop a theory-driven empirical model of 2PD. Our model allows to replicate and build on results from the screening literature on the shape of the optimal 2PD mechanism. At the same time, the model has desirable computational properties. Second, we design an experiment that helps identify the joint distribution of values in our model under two key constraints common to applications of 2PD: (i) aggregate data and (ii) small number of products/versions. Third, we implement our experimental and empirical framework in the context of flight ancillary products, demonstrating its applicability to real-world 2PD policy design settings.

Our model considers a monopolist selling two products (or, versions of a product), although the key insights can be generalized to larger numbers of products/versions. To maintain generality of applications, we remain agnostic on the nature of the differentiation between the two products; it can be quality, quantity, time of purchase, or any set of features. Motivated by theoretical results from the literature, we not only allow for heterogeneity in valuations for each product/version, but also for different degrees of co-variation between the two. More specifically, our utility specification is a model with product fixed effects: a constant and a `dummy' variable for one version. We choose the on-average-superior version as the non-omitted group (though this is not technically necessary). Here, the constant may be interpreted as the “baseline valuation,” shifting the willingness to pay for both versions; whereas the dummy represents the differentiation value. The model has two crucial features: (i) Both the constant and the dummy vary at the individual level. That is, there are random effects on both objects. (ii) The co-variance between these random effects is flexible. This co-variance determines which consumers place a higher differentiation value between the baseline and the “premium” product: those with higher or lower baseline willingness to pay. The relevance of this co-variance to the optimal 2PD policy, as will be discussed in detail, is strongly motivated by the theory literature. Last, our utility model includes an individual-product level idiosyncratic error term which serves two purposes: (i) It captures a horizontal component of differentiation between the products. This matters for  2PD because, as we will show later, the monopolist might optimally charge more for the ``on-average-inferior'' version if there is sufficient horizontal differentiation and other requirements are met. (ii) More importantly, the addition of error terms makes our model empirically estimable, unlike 2PD models from the theoretical literature on screening.

%\hl{However to fix ideas and help build intuition early on for the reader, we mention one illustrative example that will be consistent with our subsequent empirical analysis: timing of airline seat reservation.} 

%''Imagine an airline that offers two seat‑selection options: you can grab a window or aisle for \$10 at check‑in, or pay \$15 to reserve your favorite spot a week in advance. Some travelers care deeply about that early reserve (e.g., business flyers), while others just want any seat at the lowest possible fare. Moreover, the very passengers willing to pay more on average may also place a larger premium on early booking.  How should the airline set these two fees to extract the most revenue?''

%May be we can use another example, but something to fix ideas about the three constructs that follow could help a lot.

%baseline willingness to pay (how much any traveler values a seat),

%upgrade value (how much extra some care about early booking), and

%the fact that those two can move together (high value flyers might also value early booking more)

We start by using our model to deliver theoretical results which replicate and build on some of the findings established in the literature. We present formal results in section \ref{sec: theory}. Here, we provide an informal discussion assuming both versions are produced at zero marginal cost. Our analysis shows at least three quantities play key roles in shaping the monopolist’s optimal mechanism: (i) the variance across individuals in the baseline valuation (i.e., the variance of the random effect on the constant term), (ii) the covariance, across individuals, between the baseline and differentiation valuations, (iii) the relative degree of vertical vs. horizontal differentiation between the two product versions. The effect of the first factor is relatively straightforward. A higher variance in the baseline valuation means consumers are more heterogeneous. This should increase the likelihood that a monopolist seller benefits from targeting them with multiple prices, i.e., price discrimination. But we are studying a context where directly charging different prices to different consumers (i.e., first-degree PD) is not feasible. This, naturally, raises a 2PD question: should the monopolist use differential pricing between the product versions in order to implement price discrimination? This is where the other two aforementioned quantities play key roles. 

If the covariance between the baseline valuation and the differentiation valuation is positive, then charging more for the ``premium’’ product can indirectly help the monopolist charge more to those with higher baseline valuations. To illustrate, suppose that a high memory level is a feature that affluent customers of smartphones especially care about, but less affluent individuals much less so. In this case, the monopolist may optimally charge more for additional memory to target the ``high type'' customers, while also providing a low-memory version with a price targeted to the rest of the market  (even when memory level can be improved at zero marginal cost). If this covariance is negative, however, the monopolist’s optimal policy depends on the degree of horizontal differentiation between the products. First consider a case where the differentiation is primarily vertical, e.g., smartphones with different memory levels. If higher WTP customers--compared to lower-WTP ones--place a lower value on additional memory, then the monopolist no longer finds it optimal to use memory as a screening tool. It will, hence sell only the high memory version. Next, consider a substantial degree of horizontal differentiation, i.e., the on-average-inferior version is still the preferred version by a non-trivial portion of the market. In this case, perhaps counter-intuitively, we will show that the monopolist finds it optimal to charge more for the inferior product. The intuition is nuanced and is discussed in detail in section \ref{subsec: vertical and horizontal diff} along with examples. In short, the optimal selling mechanism involves selling the on-average-superior product to the majority of purchasing consumers, while pricing the other product targeted towards the minority who find it the more desirable product.

We then turn to experiment design and provide theory-driven empirical tests that help to identify the three aforementioned quantities. Formal results are provided in section \ref{sec: experiment design}.  Here, we confine to an informal discussion. First, to identify the magnitude of the variance (across consumers) in the baseline willingness to pay, we propose ``partial-market tests''.  In this test, the price of one product/version is changed exogenously, while the other price is kept fixed--i.e., the market is only ``partially'' treated. The cross-substitution rate under this test can help identify the variance of the baseline valuation. To illustrate, suppose that upon an increase in the price of high-memory smartphones, a high portion of that version’s consumers switch to the low-memory version (rather than to the outside option). This suggests that those who have a high willingness to pay for one version are likely to have higher willingness to pay for the other as well. Or, put differently, the variation in a baseline WTP (which applies to both versions) is sufficiently large to shape cross-substitution behavior.

%\hl{May be describe the two ``tests'' in words, deferring formal definitions to a later section:}

%``We show that (i) by varying only the basic seat fee across markets (we call this a partial market test as it affects only one product), one can learn how dispersed baseline values are ; and (ii) by uniformly raising prices for both options (we call this the total market test, because it affects all products), one can immediately tell whether the premium option deserves the higher price. Together, these two simple experiments let the airline or any monopolist---design a data‐driven two‑tier pricing scheme.''

%In Section ---, we introduce a parsimonious two‑version utility model that captures exactly these three ingredients (baseline values, upgrade values, and their covariance) and show how our two field tests map one‑for‑one into its structural parameters.

%\hl{Should we call this ``test'' or partial or total market ``experiment''?) SOHEIL: This makes a lot of sense. Experiment, or perhaps ``treatment'' is better than ``test''. Test makes it look like we're talking about a test statistic. I'll ask Mohammad to to change it throughout the paper.}

Our second test helps identify the covariance between the baseline valuation and the differentiation valuation. This covariance can in principle take either sign: in some markets higher baseline WTP consumers may also value the premium feature more,\footnote{A natural example of positive covariance is smartphones differentiated by memory: it is plausible that higher baseline WTP consumers (e.g., performance-oriented buyers) also place a larger premium on additional memory, which is consistent with the widespread use of memory-based versioning in this market.} while in others they may value it less.\footnote{An empirically documented case of negative covariance is \cite{gardete2025pricing}, who study news subscriptions and find a strong negative correlation ($\rho \approx -0.72$) between consumers’ willingness to pay and their intensity of usage: lower-WTP subscribers actually consume more content. Another examples is \cite{chou2024estimating} where the authors show that older customers have a higher willingness to pay for a streaming service than younger individuals; but at the same time, the older segment on average streams for a shorter amount of time.} Which case applies is therefore an empirical question, underscoring the importance of a broadly applicable test that can pin it down. The literature on estimating the covariance between random effects in discrete choice models is scant. The few studies that do examine this assume the number of products in the data is substantially larger than the number of features with random effects. However, the possible number of versions in 2PD applications tends to be few, often as small as two.  To our knowledge, our 2PD model is the first empirical setting in which the covariance between two random effects is central to the key policy of interest, rendering its credible estimation essential despite these challenges. We devise a ``total-market test'' in which the prices of both versions change by the same amount--i.e., the treatment is applied to the totality of the market. By fixing the price difference between the versions, this test is designed to shut off cross substitution. As a result, any change in the composition of the demand in response to this test can only be attributed to taste differences between marginal and infra-marginal consumers. To illustrate, consider a total-market test uniformly discounting prices across both versions of a smartphone product. Suppose in response to this treatment, the demand for the low-memory version, relative to the demand for the high-memory version, increases. This change in the demand composition implies that the marginal consumers (i.e., those responding to the discount) have a lower WTP for additional memory relative to the infra-marginal ones (i.e., those who would have purchased absent the discount). Given that marginal consumers also tend to have lower baseline valuations (since they did not purchase before the discount), this in turn implies a positive covariance between the random effects. Not only does this test work with only two products, but also it does not require data on characteristics of individual consumers. Thus, it can be applied under restrictions that mechanism designers are expected to face in real settings.

Measuring the third quantity of interest, i.e., the degree of horizontal differentiation relative to vertical, is simpler. It entails examining the demand ratio between the two versions under equal prices. If the ratio is close to zero, the differentiation is mostly vertical (this is what one would expect in the case of phones differentiated by memory levels). If it is close to one, the differentiation is mostly horizontal. This does not require performing a separate test. This is because if the prices of the products are close under the default condition (i.e., the control group), the ratio could be measured under the control group and/or the total market tests. Otherwise, it would suffice if a partial-market test equated them.

In section \ref{sec: empirical}, we set up the full empirical model to be estimated using data. Our model is a random-coefficients discrete choice model \`a la \cite{berry1995automobile} (BLP), which means that the full machinery developed for BLP estimation can be applied. At the same time, this is the very model on which we established our mechanism design results for second-degree price discrimination.\footnote{The only material difference between our empirical and theory models is the inclusion of ``structural error''. See section \ref{sec: empirical} for more details.} Thus, our framework simultaneously inherits the empirical tractability of BLP and the theoretical interpretability of screening models.

To demonstrate the applicability of our experimental-empirical framework, we implement it in collaboration with an international airline based in Asia and describe the application in section \ref{sec: application}. Our application is seat selection, and the differentiating feature between the two versions is timing of purchase. That is, seat selection can be priced differently based on whether the passenger (who has purchased the flight ticket already) books the seat within or outside a three-day interval of the date of departure. Such an exercise is a common form of 2PD, especially used in the context of products that have a set consumption date and may be booked (e.g., airfare and its ancillaries, accommodation, concerts, etc.). Under the default condition, i.e., our control group, the airline was not offering an early purchase discount. As such, the default prices of the two versions are equal. We run (i) partial-market tests offering 30\% and 60\% discounts for early purchase only, and (ii) total-market tests offering 30\% and 60\% discounts uniformly on both products. We estimate our demand model and use it to study the optimal selling mechanism. In addition, we provide a discussion on the interpretability of our experiments. We show that a combined application of our propositions on optimal mechanism and our propositions on experiment design can often help guess the optimal mechanism solely based on observing the experiment results, and before estimating the demand model and conducting counterfactual analysis.

Section \ref{sec: additional analysis} extends our main study in two directions. In Section \ref{subsec: more general empirical specification}, we implement a richer empirical specification with additional covariates, both to test the robustness of our main findings and to illustrate how the framework can be applied when more detailed data are available. In Section \ref{sec: additional simulations}, we then turn to simulated datasets that display a wider variety of patterns along three theoretically salient dimensions: (i) the variance of the baseline valuation, (ii) the covariance between baseline and differentiation valuations, and (iii) the relative weights of vertical versus horizontal differentiation. As in our main experiment, these simulations allow us to use our optimal-mechanism and experiment-design propositions to show how empirical statistics can be directly suggestive of optimal policy in a broader range of scenarios than those exhibited in our data.

Section \ref{sec: discussion} provides additional discussion on the general applicability of our framework and describes its limitations and avenues for future research. It also offers a number of practical recommendations on experiment design.

Our methodology can apply to a broad set of 2PD cases, helping mechanism designers solve for the optimal pricing policy \textit{in the field}. Underlying this applicability are three strengths of the framework: (i) The key parameters of our model are motivated by the theory of screening. (ii) Our experiment design helps to identify the key parameters--especially the covariance between baseline and differentiation valuations—under constraints that a 2PD design problem is likely to face in the field. (iii) The model can be cast as BLP, and hence, estimated efficiently. 

\section{Related Literature}\label{sec: literature}

Our paper contributes to the literature on 2PD by making a certain connection between the theoretical and empirical sides of the research on this topic. Specifically, we develop an empirical-experimental framework that allows a monopolist seller (mechanism designer) to make 2PD decisions in a theory-driven manner. Our framework achieves this goal by enabling the designer to estimate key (as suggested by theory) parameters of a 2PD model under data limitations inherent to 2PD environments.

\subsection{Theoretical Literature}

The theoretical literature on 2PD is vast. As a selling mechanism, 2PD is considered a mechanism of ``screening.'' There is a notable difference between screening and some other common selling mechanisms such as auctions. In auctions, we know since \cite{vickrey1961counterspeculation} that under conditions that are expected to hold in a range of applied settings, the optimal selling mechanism does not depend on the exact shape of the distribution of the model primitives. In fact, in a wide range of settings, the revenue-equivalence theorem holds; and different formats of auctions yield the same expected payoff for the seller. In screening, however, knowledge of this distribution is necessary to solve for the optimal mechanism. The fact that the distribution over the primitives plays a more salient role in shaping the optimal screening mechanism--than it does in determining the optimal auction--can be verified by examining most screening papers (for a non-exhaustive set of examples, see \cite{mussa1978monopoly,maskin1984monopoly,rochet2003economics,rochet2002nonlinear,adams1976commodity,mcafee1989multiproduct,armstrong1996multiproduct,armstrong2013more,armstrong2016nonlinear,desai2001quality,dana1998advance,bergemann2024unified}).\footnote{Of course, even for auctions, there are conditions about the primitives that can change the optimal mechanism (see, for instance \cite{bergemann2019revenue}). That said, the set of possible auction mechanisms is typically small (1st price, second-price, etc.). For screening, on the other hand, even a simple two-price mechanism can take infinitely many forms depending on the prices chosen for the product versions; and which of these forms is optimal depends on the distribution of model primitives.} As a result, a full implementation of a screening mechanism should involve empirical tests (or analysis of observational data) to infer that distribution. In the theory literature, however, the focus is on characterizing the optimal mechanism assuming the distribution of the primitives is known. To the best of our knowledge, the only theory paper that proposes empirical tests for inference about the primitives of a screening problem is \cite{yang2024comparison}. 

%\hl{Quality Segmentation in Spatial Markets: When Does Cannibalization Affect Product Line Design? Preyas's paper does talk about the idea of how horizontal differentiation impacts cannibalization in product line design. I think this is what he wanted us to give him a bit more credit for. }

Within the screening literature, there is a subset that is especially related to our study. This subset of the literature emphasizes that a certain feature of the joint distribution of primitives has a key role in shaping the optimal mechanism: the co-variation between (i) the baseline willingness-to-pay of a buyer--affecting all product versions--and (ii) her preference for the differentiating feature between versions. Examples of such papers--mostly applied to bundling or product-line design settings--are \cite{salant1989inducing,anderson2009price,haghpanah2021pure,ghili2023characterization,yang2021costly,yang2023nested}. Our model of 2PD allows to replicate a version of the results in these papers; which ensures the key insights they provide on optimal screening shape our empirical analysis. In addition, our empirical tests allow for the identification of the key parameters, especially the covariance parameter.

\subsection{Empirical Literature}

Empirical literature on 2PD is also vast in quantitative marketing and economics. Examples range from dynamic pricing \citep{nair2007intertemporal,hendel2013intertemporal}, to nonlinear pricing \citep{iyengar2008conjoint,iyengar2009nonlinear,iyengar2012conjoint,nevo2016usage,luo2018structural,ghili2023empirical,gardete2025pricing,chen2024empirical}, version-based pricing \citep{verboven2002quality,mcmanus2007nonlinear,aryal2021price}, share of wallet pricing \citep{mojir2022structural}, monitoring programs \citep{goldberg2021designing}, etc.

Our methodology contributes to this strand by providing a framework that enables mechanism designers to empirically implement 2PD models under constraints that are expected to be present in many applications. To this end, unlike a subset of the aforementioned papers which incorporate supply side moments into their estimation procedures (e.g., \cite{luo2018structural,aryal2021price}), we do only use demand-side moments. In addition, motivated by the theoretical literature, our empirical model keeps flexible the covariance between the baseline and differentiation values and aims to estimate it. Some of the papers in the literature impose more structure by assuming independence between the two (e.g., \cite{aryal2021price}). Finally, to estimate the parameters of interest, we design and implement empirical tests that enable identification without a requirement for a large number of products or for detailed data on consumer observables.

\subsection*{Summary}

In sum, our paper contributes to the literature by bridging between the theoretical and empirical studies of 2PD. More specifically, we develop a framework that (i) adopts a certain subset of 2PD models from the theory literature; and (ii) amends them with an experiment design that helps to empirically solve for the optimal selling mechanism under limitations likely to be faced by the mechanism designer in real-world applications. This makes our framework applicable in a wide range of 2PD settings such as quality-based discrimination, quantity-based discrimination, and discrimination based on the timing of purchase.

\section{Theoretical Analysis}\label{sec: theory}

In this section, we build a theory model of second-degree price discrimination which will motivate our empirical analysis. The objective is to build a model that (i) is parsimonious,  (ii) highlights the forces that, based on the microeconomics theory literature, play key roles in shaping price discrimination policies, and (iii) is easy to extend to an empirical model that can be estimated on data. We start with a model that allows vertical differentiation only between versions of a product. We then extend the model to allow for horizontal differentiation.

\subsection{Vertically Differentiated Products}\label{subsec: vertical diff only}

Consider two versions of a product sold by a monopolist. The ``better'' or default version is indexed 2, whereas an inferior version is indexed 1. The inferiority of version 1 comes from lack of a feature that is expected to be on average desirable by consumers. Examples of such features are: extra memory in a smartphone, higher quality seats on a flight, timing flexibility on ticket purchase, etc. The features of the two versions are exogenously given and are, hence, not subject to tuning by the monopolist.

%\hl{Sorry for not mentioning this only now, but something always bothered me about the notation, and only today, I realized it. The mixing of Greek ($\mu$, $\gamma$) and English (a,b) for parameters. I may have preferred writing}

%$u_{ik}=\alpha -\beta p_k+\mu_i+\gamma_i\times\mathds{1}_{k=2}}$

%\hl{We can go with the current notation, if we don't want to modify the proof etc at this stage. Also you have used $\alpha, \beta$ later when you relax specifications}

Consider a market in which for any individual $i$, the utility from consuming version $k$ of the product is described as follows:
 
\begin{equation}\label{eq: toy model segment 1 utility}
    u_{ik}=a-b\times p_k+\mu_i+\gamma_i\times\mathds{1}_{k=2}
\end{equation}

The utility from the outside option is normalized to zero. The term $a$ is a real-valued constant, possibly negative. The term $p_k$ represents the price of version $k$; and $b$ is the price coefficient. The term $\mu_i$ shifts individual $i$'s willingness to pay for both inside options. Given the presence of the term $a$, the term $\mu_i$  can be assumed mean-zero without loss. We assume (with loss and only in this section) that $\mu_i$ is distributed uniformly on the interval $[-\sigmamu,\sigmamu]$ where \sigmamu is a non-negative real number. The term $\gamma_i$ represents consumer $i$'s willingness to pay to upgrade to version $k=2$ from version $k=1$. We assume $\gamma_i$ is a function of $\mu_i$ given as follows:

\begin{equation}\label{eq: gamma_i mu_i}
    \gamma_i=\gamma^0+\gamma^1\times \mu_i
\end{equation}

Without loss of generality, we assume $\gamma^0\geq 0$. We assume (with loss and only in this section) that the differentiation between the two products is only vertical. That is: $\forall i: \gamma_i\geq0$, which would hold if and only if $|\gamma^1|\leq \frac{\gamma^0}{\sigmamu}$. 

The firm optimizes its profit by choosing prices $p_{1}$ and $p_2$.

Next, we deliver our results. Proposition \ref{prop: toy model} below characterizes this optimal decision making.

\begin{proposition}\label{prop: toy model}
    Consider a demand system given by equations \ref{eq: toy model segment 1 utility} and \ref{eq: gamma_i mu_i}. Suppose all marginal costs are zero. Also assume that $\mu_i\sim\text{U}[-\sigmamu,\sigmamu]$ and $|\gamma^1|\leq \frac{\gamma^0}{\sigmamu}$. Under the firm's optimal pricing policy $(p_{1}^*,p_{2}^*)$:

    \begin{itemize}
        \item The monopolist finds 2nd-degree price discrimination optimal if $\gamma^1> 0$ and $\frac{\gamma^0}{\gamma^1}<a<3\sigmamu$. In this case, the optimal prices are given by:
        
        $$p^*_{1}=\frac{a+\sigmamu}{2b},\quad\quad\quad\quad       p_{2}^*-p_{1}^*=\frac{\gamma^0+\sigmamu\gamma^1}{2b}$$
        \item In all other conditions, the firm will find it optimal to only sell version $k=2$ by setting the prices at: $$p_{1}^*\geq p_{2}^*=\frac{\sigmamu(1+\gamma^1)+\gamma^0+a}{2b}$$
    \end{itemize}

\end{proposition}

\textbf{Proof.} The proof of this proposition invokes the revelation principle and shows the above statements are true in the space of direct-revelation mechanisms. It may be found in appendix \ref{prop: toy model (proof)}. Note that even though the features of the products are assumed exogenous, 2PD decisions in this setting imply some tacit product design: by setting $p_1\geq p_2$ the monopolist will only sell version $k=2$; whereas if she sets $p_1<p_2$, she will sell both versions.

Before turning to interpretations of the two scenarios described in \cref{prop: toy model}, an observation is worth making. Our setting assumes version $k=2$ is vertically superior over version $k=1$.\footnote{Recall that this assumption will be relaxed in the empirical section.} This version is considered of (weakly) higher-quality by \textit{every} consumer in our model, while it is not more costly to produce. As a result, a socially efficient allocation of products would never allocate version $k=1$ to any consumer. What is the economic intuition, then, for the observation that the monopolist sometimes sells the inferior version? The answer is price discrimination: a profit-maximizing seller can benefit from introducing the inferior version $k=1$ and targeting its price to lower-WTP consumers, which in turn allows it to price the superior version $k=2$ with a primary focus on higher-WTP consumers. In other words, the monopolist uses the differentiating feature between the two product versions as an instrument of screening. This distortion has been studied in \cite{mussa1978monopoly,maskin1984monopoly}, and follow-up studies.

It should be noted, however, that it is not always optimal to leverage a particular product feature for price discrimination. Some recent studies \citep{anderson2009price,haghpanah2021pure,ghili2023characterization}  provide theoretical characterizations on when and to what degree product features may be used for 2PD. In fact, part of the proof of \cref{prop: toy model} leverages the main result in \cite{ghili2023characterization}.  An adaptation of that main result (which is proposed for a general bundling problem) to our setting would state the following: the monopolist would find it optimal to price discriminate if the optimal sold-alone quantity of version $k=1$ surpasses that of $k=2$. More specifically, denote by $D^*_{1}$ the volume of demand if the monopolist sells only $k=1$ and prices it optimally. Define $D^*_2$ in a similar fashion. An adaptation of this to our setting, then, holds that the monopolist should price discriminate (by selling both versions and setting $p^*_{1}<p^*_{2}$) if and only if $D^*_{1}>D^*_2$. 

With this background, we next turn to interpreting result in \cref{prop: toy model} by examining the role of each parameter in shaping the monopolist's optimal decision on whether or not to price discriminate.

\begin{itemize}
    \item \textbf{The role of \sigmamu:} A higher $\sigmamu$   makes it more likely that price discrimination is optimal. This should be expected, as $\sigmamu$ represents the magnitude of heterogeneity in consumers' WTP. The higher the consumers are differentiated in terms of their overall WTP for purchase, the more the monopolist's gain from targeting them with multiple prices.

    \item \textbf{The role of $\gamma^1$:} Although a high degree of WTP heterogeneity among consumers (captured through $\sigmamu$) means the monopolist would have a higher incentive to price discriminate, an important next question in a 2PD context is whether the particular differentiation between the two versions is a ``suitable instrument'' to implement the discrimination. This is governed by the parameter $\gamma^1$. A positive $\gamma^1$ means consumers who have a higher baseline valuation shifted by $\mu_i$, also have higher differentiation valuation shifted by $\gamma_i$. As a result, the monopolist may optimally decide to charge more for $k=2$ as an instrument for charging more to those with higher $\mu_i$. The same cannot be said when $\gamma^1<0$. In this case, a higher preference for version $k=2$ does not positively co-vary with a higher baseline willingness to pay. Hence, the differentiation between the two versions may not be used as an instrument of screening.

    To illustrate, consider a smart-phone producer which can produce phones with high and low memory levels at the same marginal cost of zero. Suppose this firm is deciding between (i) selling both high and low memory phones at different prices and (ii) selling only the high-memory version. A positive $\gamma^1$ suggests that consumers who have higher WTPs overall (for any version) happen to also have a higher preference for memory. As a result, selling two versions and charging more for memory can help the firm indirectly charge more to those higher-WTP consumers. In other words, with a positive $\gamma^1$, the firm can treat higher preference for memory as a \textit{proxy} for higher overall WTP. Or put differently, positive $\gamma^1$ means that the firm can use 2PD (based on memory levels) as a means to partially \textit{mimic a third-degree price discrimination} strategy. 

    %The situation is different under a negative $\gamma^1$: In this case, a higher preference for memory is a sign of lower WTP overall. \hl{Since this may be seen as unlikely, this is probably why we got the reaction from Miguel and Preyas that this is not important in practice. Can we think of adapting the leisure or content consumption example from there to motivate the negative correlation to reduce the negative reaction?}  As such, the strategy of charging more for additional memory does \textit{not} help the phone maker target higher WTP consumers with higher prices. Hence, selling only the better version (i.e., no 2PD) will be optimal.
    The situation is different under a negative or very small $\gamma^1$: in this case, a stronger preference for the differentiating feature is a sign of lower overall willingness to pay. As such, the feature cannot serve as a screening instrument. For example, \cite{gardete2025pricing} show that in news subscriptions, higher-WTP consumers tend to consume fewer articles, while heavier users have lower WTP. In such a case, usage-based pricing would fail as a tool for extracting surplus, and the monopolist is better off with a flat subscription fee (i.e., no discrimination on quantity). A similar intuition applies to certain ``essential'' product features in smartphones, such as basic battery capacity: consumers expect a functional battery regardless of their baseline WTP--roughtly translating to $\gamma^1\approx 0$. In both kinds of cases, the strategy of charging more for the differentiating feature does not help the seller target higher-WTP consumers. Hence, the monopolist's optimal policy is to sell only the superior version, i.e., no 2PD.

    Two additional notes: (i) Here, a positive $\gamma^1$ is equivalent to the classic Spence-Mirrlees single-crossing condition, which means our model does not impose single crossing. (ii) As the proposition shows, $\gamma^1$ impacts not only whether but also how much price discrimination takes place: $p^*_2-p^*_1$ increases with $\gamma^1$ when 2PD is optimal.

\end{itemize}

The other two parameters may be less directly related to 2PD. Hence, in order to interpret their impact on the monopolist's optimal mechanism, again, we leverage the aforementioned result from the literature on optimal sold-alone quantities: the monopolist's optimal decision involves 2PD (i.e., selling strictly positive quantities of both versions) if and only if the quantity sold of version $k=1$ if optimally priced and sold alone is larger than the same quantity for version $k=2$.

\begin{itemize}
    \item \textbf{The role of the constant $a$:} A higher constant $a$ means a higher mean for the baseline valuation of the products. This implies that the on-average-inferior version $k=1$ may be sufficiently valuable for a large enough volume of consumers to make it a worthwhile sell for the monopolist to those who may have lower differentiation valuations. Per the intuition discussed above, this makes it more likely that the monopolist finds price discrimination optimal.

    \item \textbf{The role of the average differentiation valuation $\gamma^0$:} Perhaps counter-intuitively, a higher $\gamma^0$ means the monopolist is more likely to not find it optimal to charge two different prices for the product versions. This, too, however, can be justified in light of the aforementioned intuition. A high $\gamma^0$ means that version $k=2$ is superior by a margin wide enough to incentivize the monopolist to sell this version to any consumer who would buy any version. Hence, a high $\gamma^0$ may incentivize the monopolist against price discrimination.
\end{itemize}

%\hl{We should talk about how we can present this, but the number of reference to Ghili 2023 might start to irritate reviewers and the readers. You have to think about a different way to communicate the linkages to Ghili. Find some overall place to say all the connections, but not scatter these references all over the place.}

\subsection{Vertically and Horizontally Differentiated Products}\label{subsec: vertical and horizontal diff}

Our ultimate objective is to develop an empirical framework that can be applied to real-world settings. In applications of price discrimination, products are often differentiated not only vertically, but also horizontally. This often happens because advantageous product features come not only at extra monetary expenses to consumers, but also at extra non-monetary cost. As an example, Large Language Models (LLMs) typically come in multiple, vertically differentiated models. More advanced models have better reasoning capabilities but may take longer to run. As a result, some consumers (especially in business-to-business applications) may prefer to use less advanced versions such as OpenAI's ChatGPT 4o for speed, even if each API call (i.e., instance use) is priced the same as a call on more advanced models such as ChatGPT o3. Our model, hence, would need to allow for varying degrees of vertical as well as horizontal differentiation.

There is also a more mechanical reason why we need to extend our model to horizontal differentiation. Our objective is empirical estimation. In empirical models of utility, idiosyncratic error terms are present; and error terms introduce horizontal differentiation. Utility equation \ref{eq: theory utility error terms} adds error terms to our original utility model from the previous subsection:

\begin{equation}\label{eq: theory utility error terms}
    u_{ik}=a-b\times p_k+\mu_i+\gamma_i\times\mathds{1}_{k=2}+\epsilon_{ik}
\end{equation}

Note that this modification is also accompanied by a change in the utility of the outside option. The utility of the outside option $u_{i,0}$ is now given by the idiosyncratic error term $\epsilon_{i,0}$. That is, instead of the utility itself, now only the mean is normalized to zero.

To see why \cref{eq: theory utility error terms}  introduces horizontal differentiation, observe that if the error terms are of extreme type I distribution,\footnote{Any other mean-zero and symmetric distribution with unbounded domain would work as well.} then for any price combination $(p_{1},p_2)$, there would be a positive measure of consumers who would purchase each version $k$. For other studies on how horizontal differentiation impacts optimal 2PD, see \cite{desai2001quality}.

We now turn to a key question: is the characterization in \cref{prop: toy model} robust to the addition of error terms to the utility? Note that providing a full characterization of the optimal selling mechanism in this case is difficult; because the error terms $\epsilon_{ik}$ add as many dimensions to the screening problem as the number of products plus one. We are not aware of tools in the mechanism design literature that would allow one to characterize the optimal mechanism in the presence of such idiosyncratic terms.\footnote{\cite{rochet2002nonlinear} examines a setting with random participation and do characterize the optimal mechanism. However, in their setting, randomness only impacts participation, i.e., the IR constraint; and eventual empirical implementation is not part of their objective. In our setting, the error terms impact not only the IR constraints, but also the IC, which is why direct-mechanism tools in the screening literature do not help provide a characterization in the case of \cref{prop: price discrimination error terms}.} Proposition \ref{prop: price discrimination error terms}, however, shows that a partial but useful characterization is attainable.

\begin{proposition}\label{prop: price discrimination error terms}
    Consider a demand system given by equations \ref{eq: theory utility error terms} and \ref{eq: gamma_i mu_i}. Assume that the error terms have an extreme type I distribution, and that marginal costs are zero. Also assume $\sigmamu>0$ and $\gamma^1>-1$. The following are true about the monopolist's optimal prices $(p^*_{1},p^*_2)$:
    \begin{itemize}
        \item If $\gamma^1<0$: then $p^*_{1}>p^*_2$.
        \item If $\gamma^1>0$: then $p^*_{1}<p^*_2$.
    \end{itemize}
\end{proposition}

The proof may be found in appendix
\ref{prop: price discrimination error terms (proof)} . In words, this result indicates that the characterization in \cref{prop: toy model} is not robust to introducing horizontal differentiation. In fact \cref{cor: error terms} below shows this non-robustness is the case even under arbitrarily negligible degrees of horizontal differentiation.

\begin{corollary}\label{cor: error terms}
    Proposition \ref{prop: price discrimination error terms} continues to hold if the variance of the error terms is set to any positive real number, instead of the standard level of 1.
\end{corollary}

The proof of this corollary is immediate and is based on the observation that the variance of the error terms is a normalization.

Note that the simplicity of the conditions in \cref{prop: price discrimination error terms} (i.e., the result that only $\gamma^1$ matters) is not due to the simplicity of the model. In fact the setting of \cref{prop: price discrimination error terms} is substantially more general than that in \cref{prop: toy model}: the addition of error terms, no functional form assumption  on the distribution of $\mu_i$, and no restriction on absolute value of $\gamma^1$. Yet the result is much simpler. This raises two questions: (i) what is the intuition behind the result in \cref{prop: price discrimination error terms}? (ii) Does \cref{prop: price discrimination error terms} render the result in \cref{prop: toy model} irrelevant to empirical analysis of price discrimination?

The result in \cref{prop: price discrimination error terms} and its corollary are counter-intuitive: The monopolist will charge more for $k=1$ under $\gamma^1<0$, even if horizontal differentiation (measured by the variance of the error term) is arbitrarily small and the average vertical differentiation in favor of $k=2$ (measured by the magnitude of $\gamma^0>0$) is arbitrarily large. In short, optimal pricing under $\gamma^1<0$ charges more for the ``worse'' product even if it is worse by a wide margin. To see why, fix any pricing policy $(p_{1},p_2)$ and consider two possible levels $\mu^1>\mu^2$ for the term $\mu_i$. Denote by $s_k(p_{1},p_2|\mu^1)$ the share of individuals $i$ with $\mu_i=\mu^1$ who purchase version $k$ under prices $(p_{1},p_2)$. Consider similar definitions for $s_k(p_{1},p_2|\mu^2)$. One can use simple logit demand formulas to show that under $\gamma^1<0$:

$$\frac{s_1(p_{1},p_2|\mu^1)}{s_2(p_{2},p_2|\mu^1)}>\frac{s_1(p_{1},p_2|\mu^2)}{s_2(p_{2},p_2|\mu^2)}$$

That is, as we examine higher WTP consumers (i.e., $\mu^1$ as compared to $\mu^2$), a higher share of those who purchase any version choose the ``inferior'' version $k=1$. As a result, the population of those who purchase $k=1$ has higher $\mu_i$ levels than those who purchase $k=2$ in the monotone-likelihood ratio (MLR) sense. This is what makes charging more for $k=1$ profitable. Note that this remains true regardless of how large $\gamma^0$ is.  A large $\gamma^0$ would imply that for all $\mu_i$, consumer $i$ is more likely than not to prefer $k=2$ over $k=1$. Whereas a negative $\gamma^1$ implies that those who prefer $k=1$ over $k=2$ tend to have higher $\mu_i$ than those who have the opposite preference. Both of these comparative statements can be true at the same time; but it is only the latter that shapes the ordinal comparison between $p^*_{1}$ and $p^*_2$.

For illustration, consider the aforementioned example on OpenAI's GPT-4o and o3 models where the product is an API call to the model. Suppose that the majority of the consumers prefer o3 which is a stronger model (i.e., $\gamma^0>0$ if we index GPT-4o by $k=1$ and o3 by $k=2$). Also assume, however, that the least price sensitive customers (those with highest $\mu_i$) are those financial institutions which care about speed and, hence, prefer GPT-4o over o3, suggesting $\gamma^1<0$. In this environment, it would be optimal for the OpenAI to price API calls to GPT-4o above those to o3, targeting the WTPs of financial institutions using the language model they prefer.

Do \cref{prop: price discrimination error terms} and \cref{cor: error terms} render the insights of \cref{prop: toy model} irrelevant to empirical applications of screening and price discrimination? Our answer is no. As our simulation analyses in the empirical section of the paper will demonstrate, the insights from \cref{prop: toy model} continue to ``qualitatively'' hold when the differentiation between the versions is ``mostly'' vertical.\footnote{This can happen if the standard deviation of the error term is sufficiently small. In a world where this standard deviation is normalized to 1, the equivalent condition would be for the other coefficients (especially $\gamma_0$) to be sufficiently large.} In those cases, a negative $\gamma^1$ would technically imply  $p^*_{1}>p^*_2$, but this optimal pricing policy performs only negligibly better than a uniform pricing policy that forces $p_2=p_{1}$ and chooses this common price optimally. The case of $\gamma^1>0$ is different: in markets where the differentiation between the versions is mostly vertical, we have $p^*_{1}<p^*_2$ as both propositions \ref{prop: toy model} and \ref{prop: price discrimination error terms} predict; and the respective performance improvement upon uniform pricing is non-negligible.

To illustrate, let us revisit the example of smartphones with high and low memory levels. The differentiation between these two models is vertical. That is, under $p_1\geq p_2$ one should expect little to no demand for the low memory version $k=1$. As a result,  if technically $p^*_1>p^*_2$, it will be approximately optimal to only sell version $k=2$ at $p^*_2$.

Based on this discussion, we do not see \cref{prop: price discrimination error terms} as an invalidation of \cref{prop: toy model}. Rather, we view it as a signal that a useful empirical model of 2nd-degree price discrimination should not confine to importing insights from the mechanism design and screening literature which, for tractability purposes, often restrict attention to vertically differentiated products.\footnote{To the best of our knowledge, \cite{rochet2002nonlinear} is the only paper that attempts to bring anything akin to error terms into the analysis of a multi-dimensional screening model. Even this paper, however, keeps the term fixed at the level of the individual and does not interact it with quality/quantity.} %\hl{Have you considered the Simon Anderson book treatments with logit models in theory papers? My paper with Jiwoong tries to address this stochasticity, though not using error terms}} 

We finish this section with some additional notes:

\paragraph{On the magnitude of the price difference.}  
Proposition \ref{prop: price discrimination error terms} provides only a partial characterization: it pins down the \emph{sign} of the optimal price gap $p^*_2-p^*_1$, but not its magnitude. We conjecture, however, that the size of this gap is monotone in $\gamma^1$. This conjecture is consistent with Proposition \ref{prop: toy model}, where the closed-form solution for the vertical-differentiation case shows that $p^*_2-p^*_1$ increases in $\gamma^1$. Thus, while our formal result speaks only to the sign, the conjecture suggests that the framework can also guide empirical analysis about magnitudes. Establishing such magnitude results in the presence of horizontal differentiation is challenging, as it may require new tools in multidimensional mechanism design. We leave formal proofs to future research, while here focusing on the development of an empirically implementable framework.

\paragraph{On relation to single crossing.}
In the vertical-only benchmark, the Spence--Mirrlees single-crossing condition amounts to requiring that the relative utility of version 2 compared to version 1 increases with type $\mu_i$. This holds precisely when $\gamma^1>0$, and it delivers the familiar implication that higher types sort monotonically into higher-quality allocations. Once idiosyncratic shocks $\epsilon_{ik}$ are introduced, this monotonicity no longer holds point by point: lower-$\mu_i$ consumers may sometimes prefer the higher-quality version due to shocks. Still, a \emph{stochastic} analogue survives: in expectation, or equivalently in the odds ratio of choosing version 2 over version 1 under logit errors, the choice probability shifts monotonically with $\mu_i$ if and only if $\gamma^1>0$. Proposition~\ref{prop: price discrimination error terms} can thus be read as a policy-level analogue of single crossing: the optimal price ranking mirrors the direction of type sorting, even though the strong monotonicity and IC simplifications of the deterministic case no longer obtain.

\paragraph{On marginal costs.} Before turning to experiment design, it is worth discussing how our results on the shape of the optimal mechanism generalize to the case where instead of zero marginal costs, the two product versions have marginal costs of $c_{1}$ and $c_{2}$ respectively. Proposition \ref{prop: toy model} should extend in a straightforward fashion given that its proof mainly leverages results from the literature, which allow for non-zero and heterogeneous marginal costs. Proposition \ref{prop: price discrimination error terms} also extends naturally. One can show that the proposition continues to hold if we replace the prices by the margins, comparing $p^*_2-c_{2}$ to $p^*_{1}-c_{1}$ instead of comparing $p^*_2$ to $p^*_{1}$. Given that the equivalent of this result is not proven in a previous paper, we provide it in \cref{apx: marginal costs}. In the main text, however, we focus on the case with zero marginal costs.

\section{Experiment Design}\label{sec: experiment design}

We now turn to designing an experiment to empirically estimate our model. The first step is to ask what the ideal dataset is for estimating our demand model. The answer is a dataset in which consumers are exposed, at random, to all relevant pairs $(p_{1},p_2)$, so that demand may be measured for all such combinations. Such ideal experiments are, however, likely infeasible in applications. Thus, it would be beneficial to provide minimal empirical tests that can help identify the key parameters of the model at the smallest possible cost to the mechanism designer. We turn to this task in this section.

\subsection{Theoretical Results}

The analyses in the previous subsections suggest there are three important factors that can shape 2nd-degree price discrimination policies: (i) the degree of heterogeneity in the ``baseline'' WTP among consumers, (ii) the degree to which the preference for an upgrade to the higher quality versions co-varies with the baseline WTP, and (iii) the relative strengths of vertical and horizontal differentiation between the versions in the market. In this section, we ask what are the minimal price treatments that can help identify these parameters using \textit{aggregate data} on demand.

\subsubsection{``Partial Market Tests'' and the identification of \sigmamu with aggregate data}

\begin{definition}\label{def: cross substitution}
Consider a treatment that changes $P=(p_{1},p_2)$ to $\tilde{P}$ by increasing $p_k$ to $\tilde{p}_k=p_k+r$ but keeping $p_{k'}$ fixed: $\tilde{p}_{k'}=p_{k'}$.   Define the \textbf{cross substitution rate} $\psi_{k,k'}(r)$ as: $$\psi_{k,k'}:=\frac{s_{k'}(\Tilde{P})-s_{k'}({P})}{s_{k}({P})-s_{k}(\Tilde{P})}$$
\end{definition}

In words, $\psi_{k,k'}$ shows what portion of consumers who abandon product $k$ in response to an increase in $p_k$ are those who substitute to $k'$ (rather than to the outside option). This interpretation assumes $r>0$, but this is not a necessary assumption.\footnote{For $r<0$, the interpretation of $\psi_{k,k'}$ is the share of additional demand for $k$ that is coming out of demand of $k'$ (rather than out of non-purchasers). }

Now we introduce a proposition that makes a connection between the parameter \sigmamu and a data pattern that can be directly observed under a partial-market test (i.e., a test that fixes the price of one product version and varies the other). 

\begin{proposition}\label{prop: substitution rates} \textbf{Partial Market Test:}
    Under $\sigma_\mu=0$, we have $\psi_{k,k'}=\frac{s_{k'}}{1-s_k}$. Under $\sigmamu>0$, we have $\psi_{k,k'}>\frac{s_{k'}}{1-s_k}$.
\end{proposition}

The intuition is the same as what can be found in the literature on nested and random-coefficient logit models: a high \sigmamu implies a high degree of ``vertical'' heterogeneity among consumers in terms of their overall baseline willingness to pay, which affects preference for either version relative to the outside option. As a result, knowing that a consumer $i$ has purchased version $k$ makes it more likely (than under a low/negligible \sigmamu) that this consumer would switch to the other version, as opposed to the outside option, upon an increase in $p_k$. This suggests that the substitution rate $\psi_{k,k'}$ may be useful in empirically identifying \sigmamu. 

Due to this similarity to the literature, we skip a formal proof but provide an illustration: suppose that upon an increase in the price for the high-memory version of a smartphone, some of its demand is lost; and 90\% of the lost demand switches over to the low memory version. This high cross-substitution rate suggests that those who value one version are also likely to value the other version. This, in turn, suggests that the variation in the baseline WTP for phones, which affects both versions, should be high. And that variation is governed by $\sigma_\mu$.

\subsubsection{``Total Market Tests'' and the identification of $\gamma^1$ with aggregate data}

Our utility model, described in \cref{eq: theory utility error terms}, can be interpreted as a random-coefficients discrete-choice model with two features: (i) it has product fixed effects, (ii) it has random effects on the constant term and on the fixed effect for product $k=2$, and the two are correlated. The parameter $\gamma^1$, which according to \cref{prop: price discrimination error terms}, plays a key role in shaping the optimal mechanism, is proportional to the covariance between the two random effects.

To the best of our knowledge, the literature is sparse on discussions of what variation identifies correlated random effects with aggregate data.\footnote{Note that firms often have some individual-level data on customers. However, such data rarely includes key variables (e.g., income) that are expected to underlie the particular co-variance we are interested in. As a result, we treat this identification problem as one with aggregate data. Clearly, the addition of individual-level data on measures such as income would only strengthen the estimation.} The most closely related reference is \cite{gandhi2019measuring}. The analysis by \cite{gandhi2019measuring} suggests that characteristics should independently vary among products in the data, for the covariances among their respective random effects to be identified. This, in turn, suggests that a large number of products may be needed to identify the covariances. More specifically,  when estimating a model with independent random coefficients on four variables, \cite{gandhi2019measuring} simulate a market with 15 products; but when turning to an analysis of correlated random effects, they increase the number of products to 50.\footnote{The authors do not explicitly discuss whether they independently draw product characteristics for each of their 100 simulated markets. However, based on their Appendix B2, it appears that the characteristics are indeed redrawn for each simulated market. This raises the total number of products (among which feature level may independently vary) to 5000. } %See appendix ... for a more detailed analysis on how a large number of products assists with the identification of random effects covariances with aggregate data.

Although the number of products is typically tens or hundreds in applications of random coefficients models to oligopolistic competition,\footnote{For instance, the number of products was close to 1000 in \cite{berry1995automobile}, and more than 20 in \cite{nevo2001measuring}. Neither of these papers estimates/reports off-diagonal terms in the variance-covariance matrix for the random effects.} the same is typically not the case in applications to second-degree price discrimination. In 2PD applications, the number of products, across all markets, is often as small as two, rendering the identification of the random-effects covariance challenging.  Yet, as suggested by \cref{prop: toy model} and \cref{prop: price discrimination error terms}, the covariance between the random effect on the constant and that on the fixed effect for product $k=2$ is essential to the shape of the optimal mechanism. This calls for an alternative approach to identifying this covariance in 2PD cases. 

Our analysis in this section proposes a simple test which leads to a statistic that can be directly tied to the parameter $\gamma^1$. We develop this test in the two-product environment of our model, though it can be extended to more products as well.

\begin{definition}\label{def: differential response rates}
Consider a uniform price change across products. Formally, fix a vector $P=(p_1,p_2)$ of prices and consider price vectors $\Tilde{P}=P+r$ where $r$ is a scalar. Define the \textbf{differential response rate} $\phi_{k,k'}(r)$ as: $$\phi_{k,k'}:=\frac{s_{k}(P)}{s_{k'}({P})}\mathrel{\Big/}\frac{s_k(\Tilde{P})}{s_{k'}(\Tilde{P})}$$
\end{definition}

Intuitively, $\phi_{k,k'}$ shows how the semi-elasticities of demand compare to each other across products $k$, when all of them are subject to the same level of price change, i.e., a price change that induces no cross-substitution. If $\phi_{k,k'}$ is larger than 1, it means that demand for product $k$ is more responsive than demand for $k'$ to uniform price changes.

Next we introduce a proposition that makes a connection between the sign of the parameter $\gamma^1$ and a data pattern that can be directly observed under a total-market test (i.e., a test that uniformly varies all prices, shutting off cross-substitution). 

\begin{proposition}\label{prop: differential response rates} \textbf{(Total Market Test:)}
    Assume that the dataset is generated by our model and that $\gamma^1>-1$. Then, for any vector $P$ of initial prices and any positive price change $r$, we have:

\[
\phi_{1,2}\, {\underset{(<)}{>}} \,1 \Rightarrow \gamma^1\, {\underset{(<)}{>}}\, 0
\]
\end{proposition}

The proof may be found in Appendix  \ref{prop: total market test (proof)}. The intuition is as follows: Given that a uniform price increase/decrease shuts off cross substitution, a change in the demand ratio between the two products in response to such a treatment can only be attributed to taste differences between marginal and infra-marginal buyers. For illustration, consider a uniform price increase for all smartphones (high and low memory) by a producer. Assume that upon this price increase, the proportion of customers who (conditional on purchase) buy the high-memory version increases. This suggests that those customers who care more about memory are less likely to drop out of the market in response to a price hike. Put formally, this means those with higher $\gamma_i$ are expected to also have higher $\mu_i$. In other words: $\gamma^1>0$. A similar argument would hold for a negative $\gamma^1$. This difference-in-difference type of identification argument shares similarities with arguments describing how individual-level data may be leveraged to identify the nesting parameter in nested logit models. See \cite{haile2023nonparametric}\footnote{\cite{haile2023nonparametric} is an older version of \cite{berry2024nonparametric}.} for more details.%\footnote{We thank Steven Berry for pointing to this analogy.}

To recap, as \cref{def: differential response rates} and \cref{prop: differential response rates} show, there is a simple test (i.e., uniformly changing all prices) that can generate a statistic which directly informs the sign of $\gamma^1$ regardless of the values of the other parameters. Importantly, this test does not require a large number of products.

Note that the combination of \cref{prop: price discrimination error terms,prop: differential response rates} illustrates the versatility of our framework in connecting data patterns to optimal mechanisms, even before a model of demand is estimated. Corollary \ref{cor: data-->mechanism} formalizes this:

\begin{corollary}\label{cor: data-->mechanism}

Suppose the conditions in \cref{prop: price discrimination error terms,prop: differential response rates} hold. Then, we have:
    \[
\phi_{1,2}\, {\underset{(<)}{>}} \,1 \Rightarrow p^*_2\, {\underset{(<)}{>}}\, p^*_{1}
\]
\end{corollary}

As this corollary shows, one can learn about the optimal pricing mechanism from the directly observable differential response rate $\phi$ under a total market test, without estimating a structural model of demand. Of course, the extent of such learning is limited; and to quantify the exact levels of the two optimal prices, an estimated model of demand would be required. Note that this corollary, too, will extend in a straightforward fashion to the case where each version $k$ has a marginal cost of $c_k$: the only modification will be replacing the optimal prices $p^*_k$ with optimal margins $p^*_k-c_k$ in the formula above. For more details see \cref{prop: marginal costs} and \cref{cor: data-->mechanism; MC} from \cref{apx: marginal costs}.

\subsubsection{The identification of the other parameters}

Last, we turn to the identification of other parameters, especially the vertical-differentiation parameter $\gamma^0$.

\begin{proposition}\label{prop: vertical diff}
    Consider a uniform vector of prices: $\forall k,k': p_k=p_{k'}=p\in\mathbb{R}$. Then, the following hold about the demands ratio $\frac{s_{1}}{s_{2}}$:  (i) $\partial \frac{s_{1}}{s_2} \mathrel{\big/} \partial \gamma^0<0$; (ii) $\lim_{\gamma^0\rightarrow+\infty} \frac{s_{1}}{s_2}=0$; and  (iii) if $\gamma^0=0$, then $\lim_{p\rightarrow-\infty}\frac{s_{1}}{s_{2}}=1$.
\end{proposition}

The proof of these results is straightforward and is, hence, left to the reader. The intuitive content is that the demand ratio $\frac{s_{1}}{s_2}$ under uniform prices is a good metric for assessing how much vertical differentiation there is between product versions. As item (iii) in the proposition suggests, if there are multiple vectors of uniform prices $p_k=p\, \forall k$ in the data,  the lowest-price vector is the preferred one for examining $\frac{s_{1}}{s_2}$ to gain intuition on the extent of vertical vs. horizontal differentiation.

For illustration, one would expect $\frac{s_1}{s_2}$ to be close to zero under $p_1=p_2$, when $k=1$ represents a low-memory smartphone and $k=2$ represents a high-memory version. That is, differentiation by memory level is a vertical form of differentiation. Other contexts might be different, however. For example, consider a ticket (for a flight, seat selection, concert, etc) that is differentiated based on time of purchase: $k=1$ represents early purchase and $k=2$ represents late purchase. In such a case (as we will observe in the empirical-application section of the paper), it is possible that under equal prices, $\frac{s_1}{s_2}$ takes a non-trivial value.

In sum, propositions \ref{prop: substitution rates} through \ref{prop: vertical diff} help bridge between our theoretical analysis in the previous subsections and experiment design. More specifically, subsections \ref{subsec: vertical diff only} and \ref{subsec: vertical and horizontal diff} point to three key quantities that need to be considered when studying second-degree price discrimination: (i) \sigmamu: vertical heterogeneity among consumers in their baseline willingness to pay, (ii) $\gamma^1$: the co-variation between baseline WTP $\mu_i$ and the differential WTP $\gamma_i$, and (iii) the level $\gamma^0$ of average vertical product differentiation. The results in this section suggest that to identify these key parameters, one could leverage data on (i) cross substitution rates $\psi_{1,2}$, (ii) differential response rates $\phi_{1,2}$, and (iii) shares ratio $\frac{s_{1}}{s_2}$. 

Similar to the theoretical section above, our propositions here provide only partial characterizations---often about the \emph{sign} of an effect or whether it is strictly positive. We conjecture that stronger results could be established about \emph{magnitudes}. For example, \textit{ceteris paribus}, the farther the differential response rate $\phi$ deviates from 1, the larger the absolute value of $\gamma^1$. Thus, while the formal results here speak only to signs, the conjecture suggests that the framework can also guide data-driven decision making about magnitudes. Proving such results formally lies beyond the scope of this paper. Our focus here is on building an empirically implementable framework, while leaving these theoretical extensions to future research.

We next turn to an experiment design based on the results above.

\subsection{Illustration of the design}

Inspired by \cref{prop: substitution rates,prop: differential response rates,prop: vertical diff}, our experiment design consists of two types of testing: \textbf{total-market tests} and \textbf{partial-market tests}. A total-market test, as described previously, is one in which all versions of the product are subject to the same level of price change. A partial-market test, on the other hand, subjects only one version of the product to a price modification. %Ideally, these tests should be carried out in a randomized control trial (RCT) setting at the level of individual consumers.

\begin{figure}
    \centering
    \includegraphics[width=0.5\linewidth]{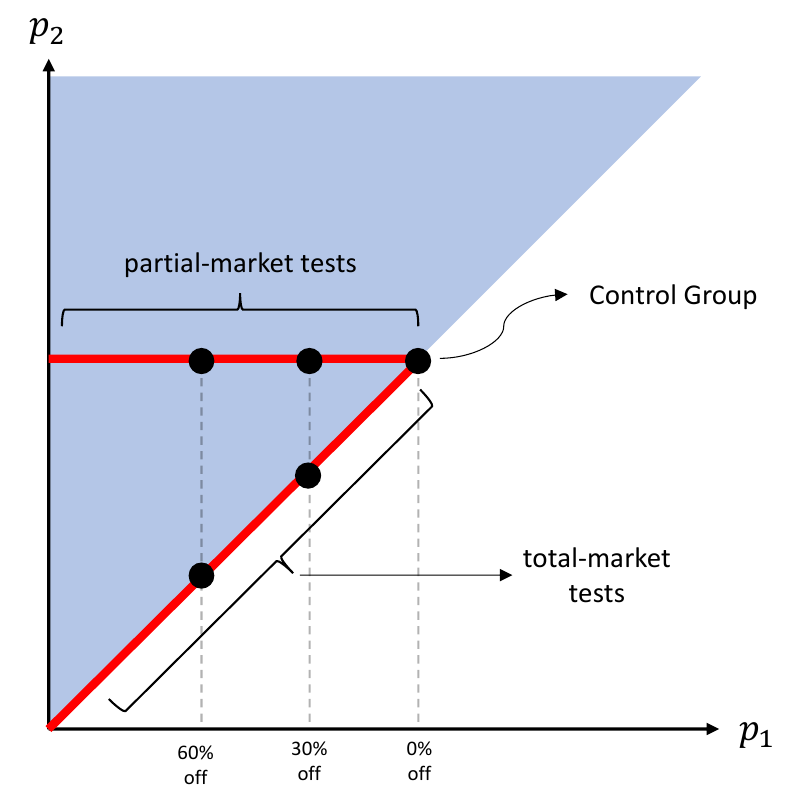}
    \caption{Schematic representation of the experiment design. On the two-dimensional space of product prices, the horizontal line captures partial-market tests, while the 45-degree line has total-market tests.}
    \label{fig:experiment design}
\end{figure}

Figure \ref{fig:experiment design} schematically illustrates the experiment design. This figure shows the possible pricing policies  $(p_{1},p_2)$ in a two-dimensional space. Each black dot corresponds to one price combination. One of the black dots shows the control group. Price combinations on the horizontal red line to the left of the control group implement partial-market tests. Price combinations on the 45-degree line departing from the control group implement total-market tests.

There are a number of features that are specific to the illustration in Figure \ref{fig:experiment design} but not essential to our experiment design. First,  the default (i.e., control-group) levels of $p_{1}$ and $p_2$ are equal in the figure. This is not an essential feature of the experiment. With any price combination as the control group, total-market tests would equally discount both prices\footnote{Whether this ``equal discounting'' should be measured in relative, percentage, terms, or absolute, dollar, terms, depends on what one thinks is the best model of how price impacts utility. With a model where price enters the utility function linearly, such as our model in this paper, we propose a discount that is uniform across versions in absolute value. With a model in which it is the logarithm of price that enters the utility function linearly (such as the specification in \cite{berry1995automobile}), we propose a discount that is uniform in relative terms. Note that the absolute and relative schemes coincide if the prices of the product versions are uniform outside of the experiment (i.e., in the control group).} while partial tests would only discount one. Second, the fact that there are exactly two separate discount levels and they are at 30\% and 60\% are not essential and may be modified to have only one, or more than two, discount levels. Third, our partial-market tests discount version $k=1$. This is not essential; and one could alternatively discount $k=2$ in applications where $p_2$ is non-trivially larger than $p_{1}$ in the control group. See \cref{subsec: experiment design considerations} for additional recommendations on how to design and implement our experiment.

With a theoretically founded experiment design at hand, we next turn to empirical analysis.

\section{Empirical Model: Random-Coefficient Nested Logit}\label{sec: empirical}

In this section, we extend the model described above to an empirical model that can be estimated using data on prices and quantities. The consumer utility equation for this empirical model is given as follows:

\begin{equation}\label{eq: utility RCNL}
    u_{imk}=\beta X_{mk}+\xi_{mk}+\mu_i+\gamma_i\mathds{1}_{k=2}-b\times p_{mk}+\epsilon_{imk}
\end{equation}

This equation closely resembles equation \ref{eq: theory utility error terms} with minor yet important modifications. First, we have an additional index $m$ for market. If an individual-level randomized control trial partitions the consumers in one market into multiple treatment groups, we use different $m$ indices for those groups. Next, instead of $a$, we have a more general term $\beta X_{mk}$, where $X_{mk}$ captures observable characteristics of product version $k$ in market $m$.  The vector $X_{mk}$ should also include a constant term. Finally, in addition to the individual-level idiosyncratic error term $\epsilon_{imk}$, we have a market level term $\xi_{mk}$ as well.

We now turn to describing the terms that, as motivated by the theoretical analysis in the previous section, capture key forces in characterizing the optimal price discrimination policy. These terms are $\mu_i$ and $\gamma_i$. The term $\gamma_i$ can be interpreted as a random coefficient on the degree of differentiation individuals see between the two versions. The term $\mu_i$ can be interpreted as a ``random coefficient on the constant term'' in the utility function. Our model of the relationship between $\gamma_i$ and $\mu_i$ follows a generalization of equation \ref{eq: gamma_i mu_i}, described below:

\begin{equation}\label{eq: gamma_i mu_i empirical}
    \gamma_i=\gamma^0Z_m+\gamma^1\times \mu_i
\end{equation}

where $\gamma^0$ is, unlike in equation \ref{eq: gamma_i mu_i}, no longer a constant. It is, rather, a vector of coefficients on market characteristics $Z_m$.

Note that this assumption implies that the correlation between $\gamma_i$ and $\mu_i$ is always within the set $\{-1,0,+1\}$. This assumption can be relaxed. See \cref{subsubsec: independent var} for more details behind this modeling choice.

Observe that this empirical model, unlike in the theoretical analysis in the previous section, we can no longer enforce the assumption that $|\gamma^1|<\frac{\gamma^0}{\sigmamu}$ given since these are parameters to be estimated from data (we did not need this assumption in proving \ref{prop: price discrimination error terms} either). If this assumption is violated by the data, there will be a second source (in addition to $\epsilon_{imk}$) of horizontal differentiation between the two versions. Also, observe that the assumption of $\gamma^0\geq 0$, which was made without loss of generality before, may not be a priori enforced here (in the form of $\gamma^0 Z_m\geq 0$) either.

Note that all the aforementioned differences between this empirical model and our theoretical version indicate that the propositions we presented need not apply to the empirical model. This is inevitable. We still maintain, however, that our propositions are useful in guiding experiment design and model interpretation. As will be discussed later in the paper, we indeed do not find empirical results to be markedly different from the predictions of our theory model. That said, it is theoretically feasible that such differences might arise in other applications.

We finish this section by emphasizing the practical properties of the empirical demand model described by equations \ref{eq: utility RCNL} and \ref{eq: gamma_i mu_i}. This model has two features. First, it closely follows the model in the previous section, hence capturing forces essential in analyzing second-degree price discrimination. Second, this is a random-coefficients logit model a la \cite{berry1995automobile} (BLP henceforth). The combination of these two features implies that the computational machinery developed in the literature for BLP models can be used to empirically estimate models of second-degree price discrimination. Perhaps most crucial among such computational methods is the contraction-mapping-based nested-loop structure first introduced by \cite{berry1994estimating} and adopted by \cite{berry1995automobile}. This means that, in the estimation process, only the heterogeneity parameters governing $\mu_i$ and $\gamma_i$ (i.e., the ``outer-loop parameters'') would need to be estimated by directly searching a multi-dimensional space. Vector  $\beta$ of parameters (i.e., ``inner-loop parameters) would be estimated through an Ordinary Least Squares regression, which can be computed in polynomial time in the number of parameters. This would allow the econometrician to specify as rich a characteristics vector $X$ as the application warrants, without concerns about the computational complexity exponentially increasing with the parameter count. To sum up: absent such a connection to BLP, the empirical estimation of models of second-degree price discrimination would suffer from a heightened risk of getting stuck at local optima (with a moderate number of parameters) or a prohibitively long computation time (with a large number of parameters).

For further clarity, note that our model can also be thought of as  Random Coefficient Nested Logit or RCNL a la \cite{grigolon2014nested}. This is because, as previously mentioned, one of our random-coefficients terms is $\mu_i$ which essentially constitutes random coefficients on the constant term. As \cite{berry1994estimating} points out, this is mathematically equivalent to allowing for correlations in error terms within two nests of inside and outside options, which is akin to a nested logit model. As a result, RCNL is technically not a generalization of BLP. Nonetheless, we find the use of the term RCNL helpful in emphasizing the nature of the random coefficients we employ.

Next, we turn to an application of this experimental-structural framework.

\section{Empirical Application: Airline Ancillary Products}\label{sec: application}

Our setting is an international airline based in Asia, which prefers to remain anonymous. Moving forward, for convenience, we often refer to it as ``the airline'' or ``the firm''. The product we study in our empirical application is seat selection. Each customer who has purchased a flight ticket may pay an extra amount in order to book a seat of their choosing as opposed to being assigned a random seat on the day of the flight. The dimension along which we vary the prices is timing: we study the firm's optimal policy on whether and to what degree it should discount booking a seat early, as opposed to closer to the departure date.

There are a few reasons why we chose the context of ancillary products, such as seat selection and purchase timing as the discrimination dimension, for the application of our econometric framework:

\begin{enumerate}
    \item Once passengers have purchased flight tickets, the market for seat selection becomes essentially monopolistic. This is useful given that our framework focuses on price discrimination under monopoly and relegates investigating the role of competition to future research.
    \item The monopoly power possessed by airlines when it comes to ancillaries makes the ancillaries a market of substantial size. By some estimates, the total ancillaries revenue of eight key U.S.-based airlines was \$4.2B/year \citep{Sorensen2023}.
    \item In spite of their large market size, ancillaries are not the main source of revenue for airlines (they trail airfare). Further, the seat assignment itself has zero additional marginal cost and as such all revenues directly contribute to profit.  Hence,  total- and partial-market tests with non-trivial discount levels (30\% and 60\%) are more feasible. These discount levels were necessary to give a degree of ex-ante assurance that the setting would help illustrate the working of our experimental framework.
    \item As will be discussed shortly, summary statistics show that under equal prices, the share of early seat purchase is consistently less than that of late  (i.e., $\frac{s_1}{s_2}$ is clearly less than 1), but the ratio is consistently non-negligible (i.e., $\frac{s_1}{s_2}$ is clearly greater than zero). This suggests both vertical and horizontal differentiation are present when it comes to purchase timing. This makes timing a useful application, given theoretical results suggesting both dimensions of differentiation play roles in shaping the prices.
\end{enumerate}

\subsection{Data and Setting}
\begin{figure}[H]
\begin{center}
    \includegraphics[width=460pt]{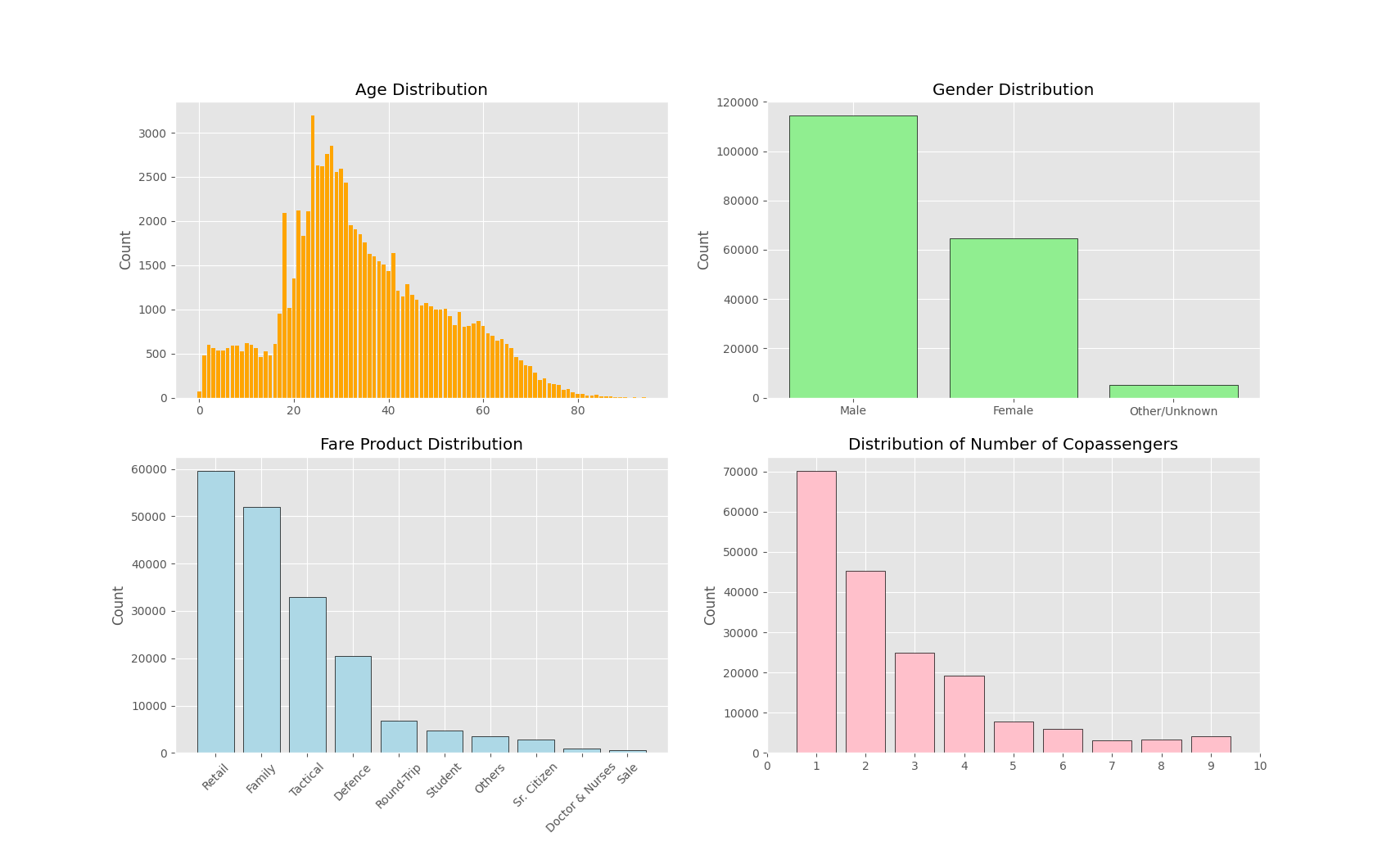}
    \caption{Demographic Data Summary (invalid ages discarded while making age bar plot)}
    \label{demosummaryfigure}
\end{center}
\end{figure}

Our data records each consumer who has obtained a flight ticket. In other words, the unit of observation is a flight-customer, or more simply, a booking. For each booking, we have information on the flight: origin and destination, length, time, and date of flight. We also have data on the passenger: gender, age, number of co-travellers, and product type. Additionally, we observe whether a seat selection was purchased by the passenger, what ``seat type'' it was if one was indeed purchased, and at what time the purchase was made. Finally, we observe the pricing of seat selection for each passenger. The pricing can indeed vary passenger by passenger because of the experiment we ran in collaboration with the airline. Figure \ref{demosummaryfigure} summarizes some statistics about the population in our data.

%[GIVE SOME DATA PATTERNS ABOUT EARLY V.S. LATE SHARE, BROADLY]

The strategic questions by the airline are two-fold. First, whether they should offer an early-purchase discount for buying seat selection at least X number of days prior to the departure? And if so, how deep should the discount be? Second, would an early purchase discount allow them to potentially raise the non-discounted, ``late purchase'' price? And by how much? In other words, the instrument of discrimination is time of purchase, with early purchase being considered product $k=1$ and late purchase product $k=2$.

%[SAY SOMETHING ABOUT THE EXPERIMENT, WHAT GROUPS WERE CHOSEN, ETC]

%Note that  our 14-days-to-departure deadline for using the discounts will be meaningless for passengers who purchase the actual flight ticket fewer than 14 days prior to departure. As a result, we divide the passengers into two groups.

The experiment was run by dividing the passengers into two sub-populations based on how long before the departure they booked their flight. The first sub-population, which we term ``D14+'' are those who purchased the flight ticket more than 14 days prior to departure. The second group, which we call ``D3-14'' are those who booked the flight between 3 and 14 days before departure. We run the experiment that was visualized in \cref{fig:experiment design} on both sub-populations. In the main text of the paper, however, we discuss only the D3-14 group; because due to higher data quality, this is the group we chose for our main specification. Table \ref{tab:D3-14 summary} depicts the early and late purchase rates by different treatment groups for D3-14.

\begin{table}[H]
\centering
\begin{tabular}{@{}lccccc@{}}
\toprule
Deadline & Discount & Count  & Total Demand & Early Demand  (k=1) & Late Demand { (k=2)} \\ \midrule
Yes      & 30\%     & 12,856 & 5.78\%          & 1.70\%         & 4.10\%          \\
Yes      & 60\%     & 12,772 & 6.36\%          & 1.90\%         & 4.40\%          \\
No       & 30\%     & 14,669 & 6.24\%          & 1.25\%         & 4.98\%          \\
No       & 60\%     & 14,615 & 8.05\%          & 1.47\%         & 6.58\%          \\
-        & 0\%      & 16,487 & 5.61\%          & 1.04\%         & 4.57\%          \\ \bottomrule
\end{tabular}
\caption{Pre- and post deadline seat selection take-up rates (aggregate across seat types) for the D3-14 group.}
\label{tab:D3-14 summary}
\end{table}

\begin{figure}[H]
    \centering
    \includegraphics[width=0.99\linewidth]{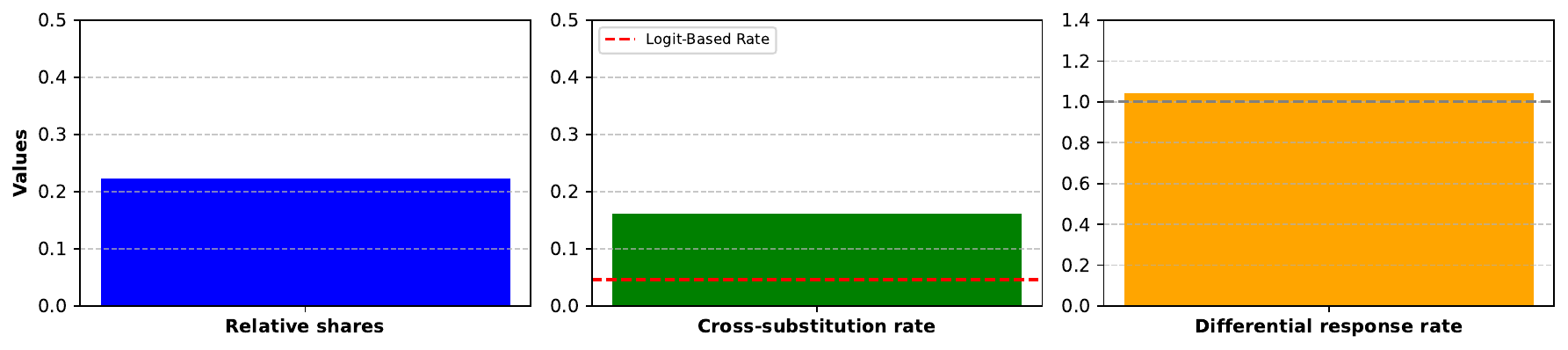}
    \caption{Visual representation of the three key summary statistics motivated by our theoretical analysis: (i) the relative shares between early and late purchase under equal prices, (ii) the cross-substitution rate under partial-market tests compared against the benchmark of ``logit cross-substitution'', and (iii) the differential-response rate under total-market tests, compared against the benchmark of 1.}
    \label{fig: summary stats data}
\end{figure}

A few patterns from these shares are worth discussing. First, and as expected, total demand increases as we offer deeper discounts or as we move from a partial discount to a total one within a fixed discount depth. The total demand (measured in percentage of booking passengers that also reserve selected seats early or late) increases from 5.61\% to 6.24\% to 8.05\% as we move from the control group to a  30\% total-market treatment and then to a  60\% one. Total demand increases from 5.78\% to 6.24\% when we move from a 30\% partial-market treatment to a 30\% total-market one. We also see a similar increase from 6.36\% to 8.05\% for 60\% treatments. 

Next, we turn to the three statistics mentioned in section \ref{sec: experiment design}: (i) the share ratio between the two product versions under equal prices, (ii) the cross substitution rates under partial-market tests, and (iii)  the differential response rates under total-market tests. All three of these statistics are visually represented in \cref{fig: summary stats data} as well.

As figure \ref{fig: summary stats data} and table \ref{tab:D3-14 summary} show, the ratio between early shares and late shares under equal prices is less than one but always non-trivial, which is suggestive that (i) on average, there is vertical differentiation between late and early in favor of late, but (ii) there is also a non-negligible degree of horizontal differentiation. The presence of horizontal differentiation is suggestive that if an estimation of our model finds a negative relationship between $\mu_i$ and $\gamma_i$ (captured by a negative $\gamma^1$), charging a higher price for early purchase may be optimal.

The average cross-substitution rate across partial-market tests, as shown by figure \ref{fig: summary stats data}, is 0.18. This is above the logit-based rate of 0.05. As suggested by \cref{prop: substitution rates}, hence, one should expect a positive variance $\sigmamu$ for $\mu_i$. By \cref{prop: toy model}, this suggests that the monopolist has an incentive to offer multiple prices and price discriminate. Whether discrimination based on time of purchase would be a suitable instrument of screening, however, is a question not possible to answer with cross-substitution rates only. We, hence, turn to the differential response rates.

%Other critical patterns about the late and early shares are less salient and do not always agree between D3-14 and D14+. Let us start from the partial-market tests. The substitution rates seem to have a wide range with a moderate-level average. For instance, for the D3-14 group, as we move from the control group to a deadlined 30\% discount, early purchase $k=1$ gains 1.70\%-1.04\%=0.66\% in share while late purchase $k=2$ loses 4.57\%-4.10\%=0.47\%, indicating a substitution rate of $\frac{0.47}{0.66}=0.71$. The rate corresponding to a comparison between the control group and a 60\% deadline, however, is 0.19. For the D14+ group, a comparison between the 30\% discount treatment and the 60\% discount treatment (both with deadlines) suggest a substitution rate, from late to early, of  $\frac{1.50-1.38}{0.82-0.63}=0.63$. For D14+, the comparison between the control group and the deadlined treatments is a bit less intuitive (this likely arises from the low overall take-up rates for the pool of consumers that were experimented on).

The differential response rate under the total market test is approximately one. That is, as we uniformly change both prices, the ratio between the demands for early and late purchase of the seat selection remains almost unchanged. By \cref{prop: differential response rates}, this suggests that $\gamma^1$ should be close to zero. This in turn, by \cref{prop: price discrimination error terms}, is suggestive that purchase timing is not a suitable instrument of discrimination, hence optimal pricing involves little to no discrimination.

\subsection{Estimation Results, and Model Fit}

Table \ref{tab:estimation results} reports the estimation results. Note that the estimated parameters are in line with what was anticipated from figure \ref{fig: summary stats data}. The parameter $\gamma^0$, i.e., the fixed effect on product $k=2$, is positive. This implies a degree of vertical differentiation between the products, consistent with the left panel of table \ref{tab:estimation results}. The parameter $\sigmamu$ is also positive and of non-trivial magnitude, consistent with the implication by the cross-substitution rate from the middle panel of the figure. This, as previously discussed, is indicative of heterogeneity in consumers' willingness to pay for seat selection. As a result, the monopolist airline would find it optimal to use the timing of purchase to price discriminate among these consumers if purchase timing is a suitable instrument of screening. By \cref{prop: price discrimination error terms}, whether timing is indeed a suitable instrument depends on the parameter $\gamma^1$. As mentioned before,  per  \cref{prop: differential response rates} and based on the right panel of figure \ref{fig: summary stats data}, the covariance parameter $\gamma^1$ is close to zero. This, as we will shortly examine, has implications for optimal pricing policy.

\begin{table}[H]
\centering
    \begin{tabular}{@{}D{.}{.}{-1}D{.}{.}{-1}D{.}{.}{-1}D{.}{.}{-1}D{.}{.}{-1}@{}}
        \toprule
             \multicolumn{1}{c}{\textbf{const}} & \multicolumn{1}{c}{\textbf{Price}} & \multicolumn{1}{c}{\textbf{\bm{$\gamma^0$}}} & \multicolumn{1}{c}{\textbf{\bm{$\sigmamu$}}} & \multicolumn{1}{c}{\textbf{\bm{$\gamma^1$}}} \\
        \midrule
         -6.086 & 0.01094 & 1.220 & 3.005 & 0.039 \\            
         (1.077) & (0.00169) & (0.804) & (0.877) & (0.243) \\
        \bottomrule
    \end{tabular}
    \caption{Estimation Results for the D3-14 group.}
    \label{tab:estimation results}
\end{table}

Table \ref{tab: model fit} shows that our estimated model generates a good model fit to the real shares. Note that, due to incorporating the structural error terms $\xi_{mk}$, our model exactly matches the data. Hence, in generating the simulated shares, we set the $\xi_{mk}$ terms to zero, in order to obtain a measure of the degree to which the ``key forces in the model'' were able to explain the shares, without help from the structural errors.

\begin{table}[H]
\centering
\begin{tabular}{@{}lccccc cc cc@{}}
\toprule
Deadline & Discount & \multicolumn{2}{c}{Total Demand} & \multicolumn{2}{c}{\begin{tabular}[c]{@{}c@{}}Early demand\\ { (k=1)} \end{tabular}} & \multicolumn{2}{c}{\begin{tabular}[c]{@{}c@{}}Late demand\\ { (k=2)} \end{tabular}} \\ 
\cmidrule(lr){3-4} \cmidrule(lr){5-6} \cmidrule(lr){7-8}
         &          & Data & Est. & Data & Est. & Data & Est. \\ \midrule
Yes      & 30\%     & 5.78\% & 5.79\%  & 1.68\% & 1.46\%  & 4.10\% & 4.33\%   \\
Yes      & 60\%     & 6.36\% & 6.11\%  & 1.93\% & 1.95\%  & 4.42\% & 4.16\%   \\
No       & 30\%     & 6.24\% & 6.65\%  & 1.25\% & 1.30\%  & 4.98\% & 5.34\%   \\
No       & 60\%     & 8.05\% & 7.90\%  & 1.47\% & 1.56\%  & 6.58\% & 6.34\%   \\
-        & 0\%      & 5.61\% & 5.55\%  & 1.04\% & 1.08\%  & 4.57\% & 4.47\%   \\ \bottomrule
\end{tabular}
\caption{Estimated vs. observed seat selection take-up rates}
\label{tab: model fit}
\end{table}

Next, we turn to analyzing the optimal pricing policy.

\subsection{Counterfactual Analysis}\label{subsec: counterfactuals}

In this section, we simulate the optimal pricing policy from the perspective of the firm. We compare the performance of that policy to two other regimes: (i) the pricing policy observed in the data, and (ii) the pricing policy that is optimal subject to $p_{1}=p_2$, meaning an optimal policy subject to no price discrimination.

Table \ref{tab: opt prices} summarizes the results. Recall that the prices are de-scaled so that they take the value of 100 under the control group. As a result of this de-scaling, the absolute values of the prices as well as those of the revenue, consumer welfare, and social welfare, are not informative. Comparisons among these numbers remains informative nonetheless.

\begin{table}[H]
\centering
\begin{tabular}{l|rrrrrrr}
\toprule
pricing & $p_{1}$ & $p_2$ & Share$_{1}$ & Share$_{2}$ & revenue & \begin{tabular}[c]{@{}l@{}}consumer\\ welfare\end{tabular} & \begin{tabular}[c]{@{}l@{}}social\\ welfare\end{tabular} \\
\midrule
\begin{tabular}[c]{@{}l@{}}Data\end{tabular} & 100.0 & 100.0 & 1.08\% & 4.47\% & 5.55 & 7.94 & 13.49 \\
\midrule
\begin{tabular}[c]{@{}l@{}}Optimal\\ Mechanism\end{tabular} & 151.9 & 155.0 & 0.78\% & 3.16\% & 6.0929 & 5.38 & 11.47 \\
\midrule
\begin{tabular}[c]{@{}l@{}}Optimal\\ Constant\end{tabular} & 154.0 & 154.0 & 0.76\% & 3.19\% & 6.0926 & 5.39 & 11.48 \\
\bottomrule
\end{tabular}
\caption{Pricing policies under three regimes: (i) observed data, (ii) optimal mechanism, (iii) optimal mechanism subject to constant pricing over time, i.e., $p_{1}=p_2$}
\label{tab: opt prices}
\end{table}

Two key lessons from the analysis. First, the optimal pricing mechanism for the monopolist raises the prices relative to the data/control group. This, unsurprisingly, increases the revenue and lowers the consumer welfare. We find the impact on social welfare to also be negative.

The second lesson, and the one more relevant to our focus, is on price discrimination. The optimal pricing mechanism sets $p^*_2>p^*_{1}$ but by a negligible margin. The ordinal comparison was to be expected from \cref{prop: price discrimination error terms} based on our estimate that $\gamma^1>0$. And the negligible margin could informally be suggested by the small magnitude of our estimated $\gamma^1$. Given that the optimal degree of price discrimination is minimal, it should also be unsurprising to observe that, as Table \ref{tab: opt prices} shows, this pricing policy does not deliver meaningfully different numbers on revenue, social welfare, and consumer welfare, relative to a scenario in which the monopolist optimizes under the constraint that $p_{1}=p_2$.

\textbf{Discussion.} We finish this application section with a brief review of how our theoretical results connect with empirical inference. Recall that \cref{prop: differential response rates} helps interpret the differential response rates from our total market test (exhibited in the right panel of figure \ref{fig: summary stats data}), suggesting that the co-variance parameter $\gamma^1$ should be close to zero. Also, recall that \cref{prop: price discrimination error terms} suggests that a small $\gamma^1$ implies a small gain from price discrimination. Putting these two propositions together, \cref{cor: data-->mechanism} suggests that a differential response rate of approximately 1 should imply negligible gains from price discrimination. This is an instance of how combining our optimal-mechanism results and demand-response results can provide insights into optimal policy direct from the data, even when a formal demand model has not been estimated. 

\section{Additional Analysis}\label{sec: additional analysis}

This section provides additional analysis that should help illustrate applications of our framework. Section \ref{subsec: more general empirical specification} introduces an empirical specification more general than the core specification used in our main analysis. Section \ref{sec: additional simulations} provides simulations of additional scenarios, further clarifying the connections between our theoretical results and empirical analysis.

\subsection{Implementing a More General Empirical Specification}\label{subsec: more general empirical specification}

Our main empirical analysis was carried out using a simple model that, except for the structural BLP error term $\xi$, exactly followed the form of the theoretical model described in \cref{sec: theory}. More specifically: (i) the term $\beta X_{mk}$ in \cref{eq: utility RCNL} is a constant $\beta X_{mk}=a$, and (ii) the term $\gamma^0Z_m$ from \cref{eq: gamma_i mu_i empirical} is  a scalar $\gamma^0$. We chose this parsimonious model due to its close alignment with the theory in \cref{sec: theory}, given that such an alignment would allow for a discussion of the close connection between Propositions \ref{prop: toy model} through \ref{prop: vertical diff} and the empirical model. Another reason for prioritizing parsimony for our main model was data related: due in part to the heavy discounts we included in our design (30\% and 60\%), the firm limited the number of individuals included in the experiment. This, in turn, makes precise identification of additional coefficients more challenging.

In this subsection, we present a more general specification with additional variables included in $\beta X_{mk}$ and $\gamma^0Z_m$. This exercise has two purposes. The first objective is to show that our substantive results from the estimation and the counterfactual analysis using the simple model are robust to including additional co-variates into the empirical specification. The second, and most important, objective is to demonstrate the applicability of the econometric framework in the general case with additional data variables.

We choose two variables and their interactions as additional co-variates in our more general specification.  One variable is the gender of the passenger booking the ticket, and the second one is whether the flight is happening during nighttime or daytime. This leads to the following two formulas for $\beta X_{mk}$ and $\gamma^0Z_m$:

$$\beta X_{mk}:=\alpha+\alpha^{N}\times \mathds{1}_{Night}+\alpha^{F}\times \mathds{1}_{Female}$$

and:

$$\gamma^0Z_m:=\gamma^{0,B}+\gamma^{0,N}\times \mathds{1}_{Night}+\gamma^{0,F}\times \mathds{1}_{Female}$$

Tables \ref{tab: general model, estimated parameters} and \ref{tab: general model, optimal prices} present the results from this empirical specification. As the tables show, there is some degree of heterogeneity in choices (both baseline and differentiation values) as a function of co-variates. Nonetheless, the overall results on the level of optimal prices and the optimal degree of 2nd-degree price discrimination remain robust.

\begin{table}[H]
\centering
\renewcommand{\arraystretch}{1.2}
\begin{minipage}[t]{0.48\textwidth}
\centering
\caption{Estimated Parameters}
\begin{tabular}{ll}
\toprule
\textbf{Inner Loop Parameters} & \textbf{Value} \\
\midrule
$\alpha$ & -4.0074 \\
Price & 0.0078 \\
$\gamma^{0,B}$ & 0.9703 \\
$\alpha^{F}$ & 0.3818 \\
$\alpha^{N}$ & -0.2567 \\
$\gamma^{0,F}$ & 0.1322 \\
$\gamma^{0,N}$ & 0.0008 \\
\midrule
\textbf{Outer Loop Parameters} & \textbf{Value} \\
\midrule
$\sigma_\mu$ & 0.8710 \\
$\gamma^1$ & 0.4150 \\
\bottomrule
\label{tab: general model, estimated parameters}
\end{tabular}
\end{minipage}
\hfill
\begin{minipage}[t]{0.48\textwidth}
\centering
\caption{Optimal Prices and Revenues}
\begin{tabular}{@{\hskip 4pt}lll@{\hskip 4pt}}
\toprule
\textbf{Prices} & \textbf{$p_1$} & \textbf{$p_2$} \\
\midrule
Constant & 1.4435 & 1.4435 \\
2PD & 1.4077 &  1.4528 \\
%3rd Degree & 1.4236, 1.3991 & \\
%           & 1.4829, 1.4505 & \\
%2\&3rd Degree & 1.4321, 1.4062 & 0.0277, 0.0243 \\
%              & 1.4931, 1.4594 & 0.0348, 0.0311 \\
\midrule
\multicolumn{2}{l}{\textbf{Revenue}} & \textbf{Value} \\
\midrule
\multicolumn{2}{l}{Constant} & 5.9704 \\
\multicolumn{2}{l}{2PD} & 5.9709 \\
%\multicolumn{2}{l}{3rd Deg.} & 5.9719 \\
%\multicolumn{2}{l}{2\&3rd Deg.} & 5.9724 \\
\multicolumn{2}{l}{} & (+0.01)\% \\
%\multicolumn{2}{l}{3rd} & +0.03\% \\
%\multicolumn{2}{l}{2\&3rd} & +0.03\% \\
\bottomrule
\label{tab: general model, optimal prices}
\end{tabular}
\end{minipage}

\end{table}

We finish this subsection with two additional discussions. First, we note that there were more observables about customers and flights in our data than what we included in the empirical specification in this section. The reason why we chose only two variables and their interactions (as opposed to incorporating more co-variates) was data limitations. The firm implemented the experiment design on a small part of the market. Thus, including more co-variates would have created small sub-markets that are too thin, thereby hindering reliable inference. That said, we examined other specifications in which other pairs of variables (as opposed to gender and time-of-day) are used. In all of those cases, the main results were robust.

The second discussion pertains to what specification we recommend for general applications. Our recommendation is that, to the extent allowed by the data, richer specifications should be preferred over the stripped-down version that resembles the theoretical analysis. Although a richer specification in our application yielded similar results as our stripped-down version, we see no guarantee that such robustness will be the case generally. As such, richer specifications are recommended when available. There is an additional reason for this: our theoretical model, in order to facilitate proofs, focused only on the covariance, among individuals, between the baseline and differentiation WTPs; and it abstracted from independent variation between those random effects. The inclusion of additional characteristics allows for such independent variation, through observables rather than random effects. In other words, the correlation between $\beta X_m+\mu_i$ and $\gamma^0 Z_m+\gamma^1\mu_i$ is no longer restricted to the set $\{-1,0,1\}$.

\subsection{Simulating Additional Scenarios}\label{sec: additional simulations}

Although the estimation and counterfactual analysis of pricing policy on the data from our application is informative of the applicability of our framework, it only illustrates a subset of possible scenarios to which the framework applies. As a result, we turn to additional counterfactual scenarios that should showcase a broader range of applications for our model. We examine a number of scenarios based on the three statistics we have emphasized throughout the paper: (i) demand ratio $\frac{s_{1}}{s_2}$ under equal prices which helps measure the degree of vertical vs. horizontal differentiation, (ii) the cross-substitution rates in response to partial market tests which sheds light on the degree \sigmamu of preference heterogeneity among consumers, and (iii) the differential response rates under total market tests which sheds light on the co-variance $\gamma^1$ between preference for the baseline version $k=1$ and the premium version $k=2$. Under each scenario constructed based on these factors, we connect key data patterns to values for parameter estimates, to implications for optimal pricing and consumer- and social-welfare.  These additional simulations should help illustrate the power of the three key statistics from our proposition in predicting outcomes of 2PD policies.

\subsubsection{Scenario 1}
\begin{figure}[H]
    \centering
    \includegraphics[width=0.99\linewidth]{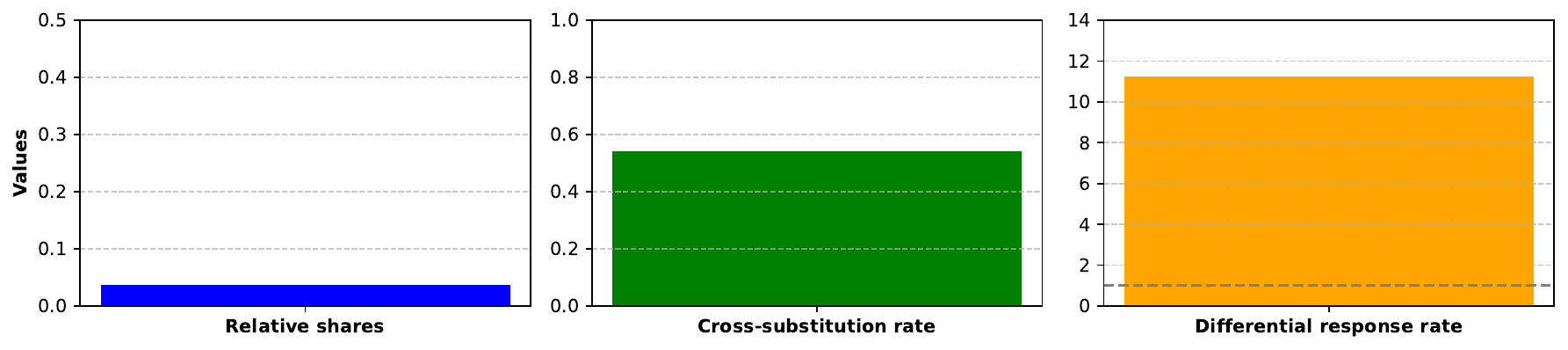}
    \caption{Relative shares, the cross substitution rate, and differential response rate, extracted from shares shown in table \ref{tab: data pattern scenario 1}. }
    \label{fig:three_panel scenario 1}
\end{figure}

Table \ref{tab: data pattern scenario 1} and figure \ref{fig:three_panel scenario 1} describe the demand levels under this scenario. There are simulated levels of demand for the control group, a partial market test of 30\%, as well as a total market test of 30\%. Table \ref{tab: data pattern scenario 1} summarizes the share levels. Figure \ref{fig:three_panel scenario 1} provides the three summary statistics discussed above.\footnote{For the shares ratio $\frac{s_{1}}{s_2}$ under equal prices, we display the shares ratio under the total market test. We make this choice although there were other options, such as $\frac{s_{1}}{s_2}$ under the control group. Our choice was guided by \cref{prop: vertical diff} which implies $\frac{s_{1}}{s_2}$ under equal prices of smaller value provides a better approximation of vertical differentiation.} 

\begin{table}[H]
\centering
\begin{tabular}{@{}lcc@{}}
\toprule\toprule
\textbf{Market}     & \textbf{Share ($k=1$)} & \textbf{Share ($k=2$)} \\
\midrule
Control Group & {0.14\%}               & {44.84\%}               \\
Partial market test ($k=1$, 30\%)            & {67.02\%}               & {8.69\%}               \\
Total market test (30\%)       & {2.96\%}               & {81.30\%}               \\
\bottomrule\bottomrule
\end{tabular}
\caption{Demand patterns representing scenario 1}
\label{tab: data pattern scenario 1}
\end{table}

Our theoretical results, combined with figure \ref{fig:three_panel scenario 1}, can help paint a picture of this market. The left panel of the figure suggests a substantial degree of vertical differentiation between the two versions $k=2$ and $k=1$ of the product. The middle panel suggests a high cross-substitution rate under a partial-market test. As is also clear from table \ref{tab: data pattern scenario 1}, when moving from the control group to the partial market test, version $k=1$ gains additional share, a significant portion of which comes out of the consumers of version $k=2$. By \cref{prop: substitution rates}, this implies some heterogeneity $\sigmamu>0$ among consumers. By \cref{prop: toy model}, such heterogeneity potentially incentivizes the monopolist to offer multiple prices and screen the consumers based on their $\mu_i$ levels. The next question is: is the differentiation between versions $k=1$ and $k=2$ a suitable ``instrument of screening''? That is, would product differentiation by offering $k=1$ and $k=2$ help the monopolist to profitably implement price discrimination? \cref{prop: differential response rates} suggests the answer would be yes if the differential response rate under a total market test is above one. The right panel of figure \ref{fig:three_panel scenario 1} confirms that this is indeed the case. This can also be verified from table \ref{tab: data pattern scenario 1}: when we move from the control group to the total market test, demand for both products $k=1$ and $k=2$ increases. Nonetheless, we observe a larger degree of relative demand increase for $k=1$. This implies that marginal consumers--those with lower $\mu_i$--tend to have a lower differential preference $\gamma_i$ for product $k=2$, relative to the infra-marginal consumers. That is, we should expect $\gamma^1>0$. In other words, preference for $k=2$ positively varies with WTP and can, hence be used as a WTP proxy and leveraged for price discrimination.

What we expect about the estimation results, based on combining our propositions with the simulated shares above, is indeed confirmed when we calibrate a model to these data. Table \ref{tab: estimated parameters scenario 1} shows that we indeed find a high $\gamma^0$, a non-trivial \sigmamu, and a positive and non-trivial $\gamma^1$.

\begin{table}[H]
\centering
    \begin{tabular}{lrr}
        \toprule
        $\gamma^0$  &        \sigmamu &  $\gamma^1$ \\
        \midrule
        4.26 & 5.0 & 0.6 \\             
        \bottomrule
    \end{tabular}
    \caption{Estimated parameters derived from shares shown in table \ref{tab: data pattern scenario 1}}
    \label{tab: estimated parameters scenario 1}
\end{table}

Finally, our simulations of optimal pricing policy also confirm what our propositions pointed to from the raw data. We find that the optimal policy charges a $p_2$ that is by about 20\% higher than $p_{1}$. This generates a 4\% higher profit relative to optimizing the prices subject to $p_{1}=p_2$. Note that a 4\% increase in profit is not a trivial number in the context of price discrimination. Especially in the case of 2nd-degree price discrimination, there are other studies that report similar numbers (e.g., \cite{ghili2023empirical}), and we are not aware of studies that report substantially higher gains. The impact of price discrimination on consumer welfare is, unlike the effect on profits, negative. But the overall effect on social welfare is positive.

\begin{table}[H]
\centering
\begin{tabular}{l|rrrrrrr}
\toprule
pricing & $p_{1}$ & $p_2$ & Share$_{1}$ & Share$_{2}$ & revenue & \begin{tabular}[c]{@{}l@{}}consumer\\ welfare\end{tabular} & \begin{tabular}[c]{@{}l@{}}social\\ welfare\end{tabular} \\
\midrule
\begin{tabular}[c]{@{}l@{}}Optimal\\ Constant\end{tabular} & 74.0      & 74.0      & 2.13\%   & 78.02\%   & 59.36 & 25.75  & 85.10                                                 \\
\midrule
\begin{tabular}[c]{@{}l@{}}Optimal\\ Mechanism\end{tabular} & 64.7      & 78.0      & 32.74\%   & 51.82\%   &  \begin{tabular}[c]{@{}l@{}}61.71\\ {\tiny (+4\%)} \end{tabular} & \begin{tabular}[c]{@{}l@{}}24.14\\ {\tiny (-6.3\%)} \end{tabular}   & \begin{tabular}[c]{@{}l@{}}85.85\\ {\tiny (+0.9\%)} \end{tabular}                                                 \\
\bottomrule
\end{tabular}
\caption{Pricing policies for scenario 1 under two regimes: (i) optimal mechanism subject to constant pricing over time, i.e., $p_{1}=p_2$, (ii) optimal mechanism}
\label{tab: pricing and welfare analysis scenario 1}
\end{table}

To summarize, this analysis with simulated data goes beyond our field experiment to further showcase the usefulness of the theoretical results, in conjunction with our experiment design, in tying empirical patterns to optimal policy, often even before a structural model is estimated.

\subsubsection{Scenario 2}

\begin{figure}[H]
    \centering
    \includegraphics[width=0.99\linewidth]{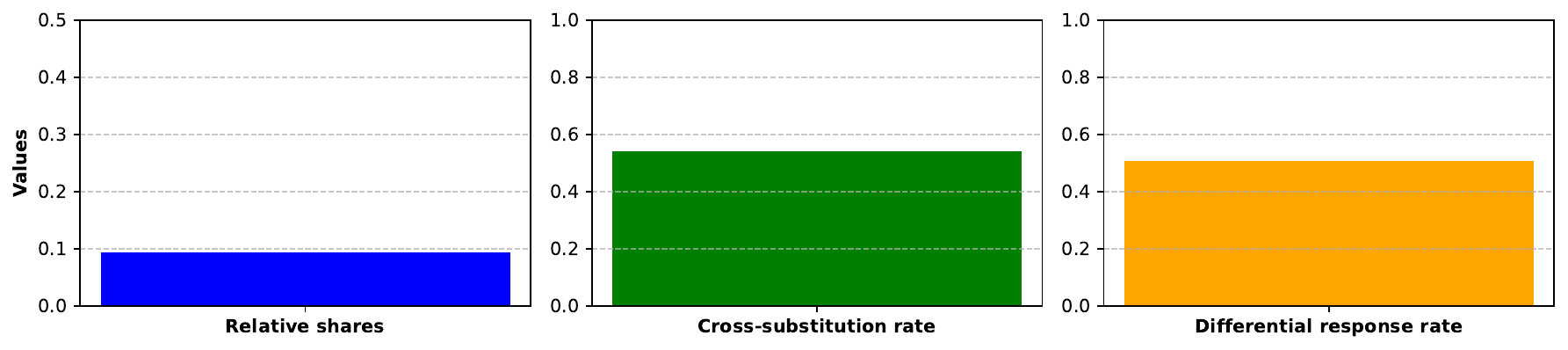}
    \caption{Relative shares, the cross substitution rate, and differential response rate, extracted from shares shown in table \ref{tab: data pattern scenario 2}. }
    \label{fig:three_panel scenario 2}
\end{figure}

This scenario is summarized in table \ref{tab: data pattern scenario 2} and figure \ref{fig:three_panel scenario 2}. These numbers portray a similar market to what we considered in the previous scenario, but with one major difference: the differential response rate shows that under a 30\% total market discount, demand for the superior version $k=2$ increases by a wider relative margin, compared to the demand for $k=1$.

\begin{table}[H]
\centering
\begin{tabular}{@{}lcc@{}}
\toprule\toprule
\textbf{Market}     & \textbf{Share ($k=1$)} & \textbf{Share ($k=2$)} \\
\midrule
Control Group & {7.67\%}               & {41.49\%}               \\
Partial market test ($k=1$, 30\%)            & {75.00\%}               & {4.95\% }              \\
Total market test (30\%)       & {8.57\%}               & {91.40\%}               \\
\bottomrule\bottomrule
\end{tabular}
\caption{Demand patterns representing scenario 2}
\label{tab: data pattern scenario 2}
\end{table}

%Similar to scenario 1, in this scenario we also expect a degree of preference heterogeneity $\sigmamu >0$. This can be inferred combining \cref{prop: substitution rates} with the high cross-substitution rate under the partial market test. This, once again, implies that the monopolist would have the incentive to use the versioning based on $k=1,2$ for price discrimination, if the differentiation between the two version provided a suitable instrument of screening. Unlike scenario 1, however, in our current scenario, this differentiation is not a good instrument. As the differential response rate, combined with \cref{prop: differential response rates}, implies, we should expect a negative $\gamma^1$. That is, higher WTP consumers tend to have a lower differentiation willingness to pay for version $k=2$.

What is the monopolist's optimal policy in this market? The high cross-substitution rate suggests a non-trivial preference heterogeneity \sigmamu according to \cref{prop: substitution rates}. In conjunction with this, \cref{prop: toy model} suggests that the monopolist would want to offer multiple prices in order to capture a higher surplus. The next question is whether the versioning based on $k=1,2$ provides a suitable means of screening and capturing surplus. \cref{prop: differential response rates}, combined with the observed differential response rates result from the right panel of figure \ref{fig:three_panel scenario 2}, suggests that the answer is no. The co-variation parameter $\gamma^1$ is expected to be negative, according to this proposition. Table \ref{tab: estimated parameters scenario 2} below confirms these expectations:

\begin{table}[H]
\centering
    \begin{tabular}{lrr}
        \toprule
        $\gamma^0$  &        \sigmamu &  $\gamma^1$ \\
        \midrule
        5.2 & 4.6 & -0.73 \\             
        \bottomrule
    \end{tabular}
    \caption{Estimated parameters derived from shares shown in table \ref{tab: data pattern scenario 2}}
    \label{tab: estimated parameters scenario 2}
\end{table}

Note that \cref{prop: toy model,prop: price discrimination error terms} may have different implications for optimal pricing in this scenario. \cref{prop: price discrimination error terms}, which directly applies to the setting we are studying, implies that $p^*_{1}>p^*_2$. \cref{prop: toy model} only allows for vertical differentiation between the two products and does not directly apply here. However, given that the differentiation between the two versions is ``approximately vertical'' (as suggested by the left panel of \cref{fig:three_panel scenario 2}), its implications might be of relevance. This proposition suggests that price discrimination is not profitable, i.e., the monopolist would not lose profit by setting $p_2=p_{1}$.  Insights from both of these propositions are reflected in table \ref{tab: pricing and welfare analysis scenario 2}. In line with \cref{prop: price discrimination error terms}, we can indeed see that $p^*_{1}>p^*_2$. However, given the large degree of vertical differentiation, the implication from \cref{prop: toy model} holds in approximation: unlike scenario 1 in which the optimal flexible policy delivered a 4\% higher profit relative to the optimal uniform-price policy, here the improvement is only 0.2\%.  This is in line with our previous conjecture that when the differentiation between the products is primarily vertical, \cref{prop: toy model} holds in approximation.

\begin{table}[H]
\centering
\begin{tabular}{l|rrrrrrr}
\toprule
pricing & $p_{1}$ & $p_2$ & Share$_{1}$ & Share$_{2}$ & revenue & \begin{tabular}[c]{@{}l@{}}consumer\\ welfare\end{tabular} & \begin{tabular}[c]{@{}l@{}}social\\ welfare\end{tabular} \\
\midrule
\begin{tabular}[c]{@{}l@{}}Optimal\\ Constant\end{tabular} & 87.0      & 87.0      & 8.58\%   & 87.24\%   & 83.43 & 13.26  & 96.69                                                 \\
\midrule
\begin{tabular}[c]{@{}l@{}}Optimal\\ Mechanism\end{tabular} & 92.22      & 87.0      & 3.62\%   & 92.21\%   & \begin{tabular}[c]{@{}l@{}}83.61\\ {\tiny (+0.2\%)} \end{tabular} & \begin{tabular}[c]{@{}l@{}}12.96\\ {\tiny (-2.3\%)} \end{tabular}   & \begin{tabular}[c]{@{}l@{}}96.57\\ {\tiny (-0.1\%)} \end{tabular}                                                 \\
\bottomrule
\end{tabular}
\caption{Pricing policies for scenario 2 under two regimes: (i) optimal mechanism subject to constant pricing over time, i.e., $p_{1}=p_2$, (ii) optimal mechanism}
\label{tab: pricing and welfare analysis scenario 2}
\end{table}

\noindent\textbf{Summary and other simulations.} Our two simulation analyses above showcase how our theoretical results, in conjunction with our proposed tests, can help understand the fundamentals of a price discrimination market even before estimating a structural model. In both cases, one can use our propositions to make conjectures about the parameter estimates and the shape of the optimal policy, and the relative performance of the optimal policy, once compared against uniform pricing.

\section{Discussion}\label{sec: discussion}

\subsection{General Applicability}
Our methodology helps to empirically address a set of screening problems with a focus on second-degree price discrimination. At the firm level, this methodology is applicable by any firm that offers multiple versions of a product and is able to carry out partial- and total-market tests. We expect digital and technology products to constitute a large subset of the possible applications. Cloud computing, LLM-based products, and digital devices such as smartphones and laptops are among such examples. In general, any setting in which a firm considers whether to ``version'' its products based on one or more features can benefit from our framework. Our total-market tests can be implemented by issuing potential consumers discount offers that can be applied regardless of the version purchased. Partial-market tests can be implemented through discount offers that apply only to one version.

In the particular case of the airline we studied, our results suggested that the price of seat selection should substantially increase (by about 50\%) and that no 2PD was necessary based on purchase timing. This was recommended to the firm and was subsequently implemented.

%[STH ABOUT HOW IT APPLIES BEYOND ONE FIRM]]

\subsection{More on Modeling Choices}
In this section, we discuss some choices made on how to set up the random effects in our utility function.

\subsubsection{Why no random coefficient on price}

Our demand model, as previously discussed, is one of random coefficients discrete choice, otherwise known as BLP. In most BLP models of demand, price is one of the coefficients on which there is random effect.  Our utility function, as described in \cref{eq: theory utility error terms} and \cref{eq: utility RCNL}, only has a fixed coefficient $-b$. There were a number of reasons behind this modeling choice, which we discuss below:

\begin{enumerate}
    \item The current model, as our theoretical results show, lends itself to an analysis of optimal 2PD decisions that can be formalized through propositions. We were unable to find equivalents to \cref{prop: toy model} and \cref{prop: price discrimination error terms}  for a model with random coefficients on prices. We believe the provision of these formal propositions connecting theory to empirical analysis is an important part of this paper's contribution. As such, we chose a model specification that made this connection possible.\footnote{Of course difficulty to obtain formal results does not imply that a model allowing for a heterogeneous price coefficient to co-vary with a differentiation coefficient is irrelevant to 2PD. Such a co-variance, and many other factors our 2PD model abstracted away from, might indeed have roles in shaping optimal 2PD decisions. We leave a thorough investigation of those to future research.}
    \item We allow for a random coefficient on the constant term. This, as discussed before, effectively creates a nested-logit structure, addressing the key drawback of standard logit models when it comes to substitution patterns, lessining the need for random coefficients on prices.
    \item More broadly, in our experience with random coefficients on prices, we have found that functional form assumptions on the probability distribution over the random coefficient have a disproportionately large role in shaping all the main outcomes.\footnote{To illustrate, consider a normal distribution on the price coefficient $b_i$ across individuals $i$. Such an assumption would imply that, unless the distribution is degenerate, there is always a positive mass of customers $i$ who strictly prefer higher prices: $b_i<0$. It immediately follows that the optimal policy for the firm is to charge infinitely high prices for its products. Note that a normal assumption on the price coefficient has been used in a number of BLP studies published in the literature. Also, even other distributional assumptions do not fully solve the problem. Consider log-normal, for example. The behavior of a log-normal distribution around zero will still have an outsized effect on all outcomes relative to the shape of the distribution away from zero.} 
    \item Finally, another important role for random coefficients on the price is when it comes to welfare analysis. Under a fixed coefficient on prices (as in our model), a transfer of cash between any pair of customers is welfare-neutral.  Under random coefficients, however, a transfer of cash from a customer with a lower $b_i$ to one with a higher $b_i$ is welfare-enhancing. As such, if the main purpose of a study on 2PD is the welfare effects and we want the welfare metric to reward ``redistribution,'' random coefficients on prices are recommended. Otherwise, we believe a model with random coefficients on the constant and on the version dummies would suffice.
\end{enumerate}

We finish this discussion by noting that our decision to exclude random coefficients on prices does not preclude their inclusion in future applications of our framework. Rather, it will require a more careful interpretation of the results.\footnote{For instance, consider a setting in which there are random effects $b_i$ on prices but the differentiation coefficient $\gamma_i$ is a constant: $\gamma_i\equiv \gamma_0$. One can show that if we only allow for a co-variance between $b_i$ and $\mu_i$ (i.e., no independent variation),  then  \cref{prop: differential response rates}  can be repurposed to make a connection between the differential response rates and the sign of the covariance. In other words, \cref{prop: differential response rates} would ``survive'' the introduction of random coefficients on prices as long as we remove random coefficients on the constant. However, it is not clear what the equivalent of \cref{prop: price discrimination error terms} will be in this environment. Moreover, if we consider a more general environment in which there are random effects on all three (i.e., the constant, the differentiation dummy, and the price), neither of the results will survive.}

\subsubsection{Why no independent variation between the random coefficients on baseline and differentiation WTPs}\label{subsubsec: independent var}

As previously mentioned, our model forces a deterministic relationship between the random coefficient $\mu_i$ on the constant and the random coefficient $\gamma_i$ on the differentiation. This, combined with the fact that the relationship is assumed linear, forces the correlation between the two to be in the set $\{-1,0,1\}$. There were two reasons for this modeling choice:

\begin{enumerate}
    \item Similar to the random coefficient on price, an independent variation between the two random effects would have made it prohibitively difficult to prove our key results \cref{prop: price discrimination error terms} and \cref{prop: differential response rates}. Put differently, we provide economic theory on why the covariance between these two random effects is tightly linked to optimal 2PD decisions; but we were unable to construct similar results or an intuitive argument connecting independent variation between the random effects to optimal 2PD decisions.
    \item We believe this assumption does not come at a significant cost. This is because our general specification, examined in \cref{subsec: more general empirical specification}, allows for this independent variation through observable characteristics. As that section showed, in the case of our application, the results are robust.
\end{enumerate}

Similar to the previous discussion, our empirical model can be readily extended to introduce independent variation between the random effects (i.e., through unobservables). In that case, our theoretical results should be used and interpreted with caution, as they no longer formally hold, but can perhaps still aid intuition.

\subsubsection{Why not a demand model with forward-looking customers?}

2PD models of products differentiated by the time of purchase are often described by models of advance pricing \citep{dana1998advance,shugan2000advance,shugan2004advance,shugan2005advance} which itself is a special case of ``sequential screening'' a la \cite{courty2000sequential}. A core component of these models is that customers in earlier periods face uncertainty about their WTP for the product. Later, this uncertainty resolves. If there is heterogeneity among customers in the degree of uncertainty faced, then pricing based on time of purchase (i.e., ``advance pricing'') can be used for 2PD. A well-known example is pricing of airfare (note the difference from seat selection) and discriminating between business and leisure travelers. Business travelers often face substantial advance uncertainty on whether they need to travel a specific route; but conditional on wanting to fly, they often have substantially higher WTPs relative to leisure travelers. Under conditions characterized by some papers in the literature, it will be optimal for a monopolist to charge higher amounts closer to the departure. This strategy implements 2PD by (i) incentivizing less uncertain but low-WTP leisure travelers to purchase early while (ii) nudging uncertain but high-WTP business travelers to wait until the resolution of their uncertainty, accepting that they pay a higher price for it.\footnote{Note that another approach to modeling time-based pricing of airfare would be one of third-degree price discrimination, 3PD. In this approach, higher WTP travelers are not modeled as having a higher degree of ex-ante uncertainty about whether they should fly.  They are, rather, modeled as ``arriving'' at the market in later periods. This 3PD modeling approach is commonly used in empirical studies of the airline industry. For instances, see \cite{lazarev2013welfare,williams2020dynamic,aryal2021price}.}'\footnote{Another 2PD approach in these environments, which generalizes time-of-purchase pricing, would be to offer (full or partial) insurance on purchase. For examples of these approaches see \cite{courty2000sequential,yang2023match}. For additional empirical work on sequential screening see \cite{miravete2002estimating}.}

Our modeling approach in this paper was \textit{not} one of sequential screening. Rather, we worked with a standard model of quality-based price discrimination. We had two reasons for this modeling choice. The first, and more important, reason is methodological. The model presented in this paper has a substantially wider range of applications relative to sequential screening. The second reason behind our choice is substantive. Unlike actual airfare, there is no reason to believe that individuals face uncertainty about their WTP for seat selection for a flight that they have already booked. As a result, we think of the time-of-purchase choices made by consumers as merely a manifestation of time preference rather than uncertainty resolution.%\footnote{\hl{Not sure if this footnote is needed. We can rewrite the discussion of sequential screening in a more straightforward way.} In fact earlier versions of this paper were titled ``An Empirical Analysis of Sequential Screening'' and were focused on uncertainty resolution. But due to a lack of empirical or anecdotal evidence for uncertainty playing a role, we switched to a simpler empirical  2PD framework with a wider range of applications.}

\subsection{Future Research, Extensions, and Limitations}

\subsubsection{Future research}

The empirical implementation of screening models is a difficult yet crucial task. This is especially the case when one's objective is to make the framework useful to the mechanism designer (e.g., monopolist), which requires avoiding the use of moments that assume optimal policy on the part of the designer. We see our methodology as an early attempt at this objective. Additional advances, both on the mechanism design side and on the econometrics side, should further help with this objective.

On the theory side, we see two key areas. The first would be to extend the models of screening in the literature to settings in which the utility function includes idiosyncratic terms at the \textit{individual-item} level. This is equivalent to the $\epsilon_{ik}$ terms in our specification. These terms help connect screening models to empirical applications because they help to rationalize a wide range of datasets which may not otherwise perfectly fit with the behavior predicted by a theoretical model. We are not aware of any study that allows for such randomness at the $ik$ in a general fashion. Our treatment of the terms $\epsilon_{ik}$ imposed an extreme type I distribution assumption. The closest other study that we are aware of is \cite{rochet2002nonlinear}, which introduces random participation, equivalent to having $\epsilon_{ik}$ only for the case where $k$ is the outside option. This structure allows the idiosyncratic term to only impact the individual rationality (IR) constraint, rather than both IR and IC.  We believe a general treatment of the idiosyncratic terms (i.e., for all $ik$, without functional form assumptions) will add significant value; but at the same time, it will complicate the screening problem, since it introduces disturbances to the shape of the incentive compatibility (IC) constraints. To illustrate this point, note that with the introduction of idiosyncratic errors, the traditional single-crossing property is unlikely to hold in most screening models in the literature, which would preclude use of most of the techniques developed for characterizing optimal mechanisms.

Another direction of future research in the theory of screening that would facilitate empirical applications is the development of results for experiment design. To the best of our knowledge, the chief focus in the theoretical literature on screening models is the characterization of optimal mechanisms under the IC and IR constraints as a function of the model primitives. To increase the likelihood of empirical applications, we believe theories could supply experiment-design results. Those are results that predict how certain patterns of buyer behavior under a pre-determined (and not necessarily optimal) mechanism can help infer certain features of the model primitives. \cref{prop: substitution rates,prop: differential response rates}  illustrate this in our specific context. We believe these types of results can add value to screening models more broadly.

On the empirical front, a key avenue for future research pertains to a ``meta'' question regarding the profitability of 2PD in the broad sense. Empirical research on 2PD (including the present paper) has consistently found profitability levels for 2PD that are at times negligible and at most in the ball-park of 10\% (for examples, see \cite{hendel2013intertemporal} for sales, \cite{iyengar2012conjoint, nevo2016usage, luo2018structural, ghili2023empirical} for nonlinear pricing, and \cite{leslie2004price,draganska2006consumer,aryal2020empirical} for feature-based discrimination).\footnote{Though not small, 10\% falls considerably short of what could often be achieved with third-degree PD (3PD). For comparison, \cite{dube2017scalable} shows that a 3PD mechanism with sufficient data on individual customers and an adequately flexible model to base the pricing on those characteristics can achieve profitability performances close to first-degree price discrimination (1PD). Given that 1PD often delivers close to twice as much in profit relative to no price discrimination (for instance, see \cite{aryal2020empirical,ghili2023empirical}), this means evidence so far points to a considerable performance gap between 2PD and 3PD. } Nonetheless, theoretically, the profitability of 2PD is not bounded at levels close to 10\%. As a result, an important empirical question is as follows: Is ``at most about 10\%'' a sound prior on the profitability of 2PD (relative to no PD) in a wide range of contexts? The answer to this question, which in our view has significant practical implications, cannot be provided by a single paper. It should, rather, emerge as additional studies measure the profitability of 2PD across a range of industries and products.

%the design of minimal experimental treatments for the identification of screening models without supply side moments. Another important direction would be extending our analysis of experimental price variation to a study of optimal instruments in an environment where prices may be endogenous. We conjecture that, in line with our tests, the suitable instruments would also need to be of two types: (i) instruments inducing variation in the differential value between the products, in a similar spirit to our partial-market tests and in line with differential-valuation  instruments a la \cite{gandhi2019measuring}; (ii) instruments inducing exogenous variation in market participation without inducing cross-substitution, in a similar spirit to our total-market tests. We conjecture that with these two types of instruments,  an empirical model of monopoly 2PD may be identified without supply-side moments and with aggregate non-experimental data. However, we leave a full analysis of this subject to future research.

Aside from the above general direction, more specific directions of future research on empirical implementation of 2PD models would entail a number of possible extensions to our model, which we turn to next.

\subsubsection{Extensions to our model}

We now discuss several extensions to the model that can be useful in empirical applications. First, instead of the linear relationship $\gamma_i\equiv\gamma^0+\gamma^1\times \mu_i$, one can extend the relationship between the two random effects to allow for nonlinearity. Such an extension would lead to different functional forms for the marginal distribution of $\mu_i$ and that of $\gamma_i$. If such a model is to be empirically estimated using partial- and total-market tests, we expect that multiple levels of treatment are necessary. In particular, we expect that the shape of the differential response rate across different total-market treatment levels will help to identify the curvature of the function $\gamma(\mu_i)$. We conjecture that all of our propositions extend to this nonlinear case in a relatively straightforward fashion. Nevertheless, we leave a formal extension to future research.

Second,  the empirical model can be readily extended to capture a full joint distribution on $\mu_i$ and $\gamma_i$, compared to the current setting in which $\gamma_i$ is modeled as a deterministic function of $\mu_i$. Such an extension will allow for the correlation between $\gamma_i$ and $\mu_i$ to assume any value in the interval $[-1,1]$ whereas our current model only allows for $\{-1,0,1\}$. Note, however, that as mentioned before, proving our theoretical results would be difficult for such an extension. Also as discussed in section \ref{subsec: more general empirical specification}, our general model does capture a flexible correlation between baseline and differentiation WTPs through observable characteristics.

A third direction in which our model can be extended would be by increasing the number of products beyond two. Such applications are relatively common, especially in cases of three versions of a product. We expect our results to naturally extend to the cases of $K>2$ products. That said, a formal analysis falls beyond the scope of this paper.

Finally, an important extension would be incorporation of oligopolistic competition and the implications it might have for firm-optimal and socially efficient screening.\footnote{For some empirical studies on price discrimination under competition, see \cite{borzekowski2009competition,aryal2020empirical}.} We expect this extension to be challenging for a number of reasons. Chief among those reasons is the presence of multiple equilibria (see \cite{ellison2005model} for more details). We note, however, that our monopolistic model of 2PD might still be a useful tool to analyze 2PD questions in oligopolistic markets under a range of market scenarios.\footnote{One obvious case where a monopolistic model is adequate is when the focal firm has a large market share and substantial market power (i.e., scenarios where the focal firm is ``approximately'' a monopolist). In addition, our monopolistic model of 2PD may be applied to oligopolistic markets as long as we expect the prices of the competitors to respond minimally to the strategy chosen by the focal firm (this could happen when the focal firm is smaller than the main players in the market by an adequately wide margin). In such cases, the prices chosen by other firms are exogenous to the strategy of the focal firm. As a result, one can assume that competing products are adequately captured under the outside option value.}

\subsection{Additional Considerations in Experiment Design}\label{subsec: experiment design considerations}

In this section, we discuss a number of practical considerations in deciding the details of the experiment design for general application. First, as Figure \ref{fig:experiment design} shows, both partial- and total-market price variations are generated by a price decrease, as opposed to a price increase, relative to the control group. This is not econometrically essential, but perhaps the only possibility from a practical standpoint. Price decreases could be framed as an ``offer'', while experimenting with price increases would face severe feasibility challenges. 

Second, we recommend at least two distinct price changes for each test. This should help alleviate potential placebo-effect concerns by creating varying degrees of treatment as opposed to just a binary treatment-control comparison. Note that in an extension of our model in which the relationship between the differentiation WTP $\gamma_i$ and the baseline $\mu_i$ is nonlinear, more than one treatment level is essential, even if placebo effects are not a concern.

%Third, if under the control group, the two prices are different, we propose that the partial-market test discount the more expensive product (i.e., version $k=2$) and choose one of the discount levels so that the two prices are equal. Per \cref{prop: vertical diff}, this should allow for an easier measurement of the degree of horizontal vs. vertical differentiation. 

The third is regarding the choice of the right product version for the partial-market test. If under the control group we have $p_1=p_2$, we recommend that the ``on average inferior'' product (i.e., the one having a smaller market share under the control group) to be chosen for the partial market test. This is what we did in our application. If, however, under the control group we have $p_1<p_2$, we recommend using $k=2$ for the partial market test. We also recommend using a discount level that sets $p_1=p_2$ for at least one treatment group. This will produce interpretable $\frac{s_1}{s_2}$ data points, which might help with managerial decisions even before estimating a structural model.

Our fourth practical observation pertains to the choice of the discount depth in designing the treatments. Our experience suggests that choosing more intensive treatments, such as 30\% or 60\% discounts, has an advantage and a disadvantage. The advantage is straightforward: the effect size of an intensive treatment is typically larger and, hence, less expected to be masked by noise or by inherent imperfections of field experiments (such as placebo effects, etc.) The disadvatage is that large treatment sizes are more likely to push firms towards limiting the pool of the population on whom the experiment is run. This might hinder the researchers' ability to include as many control variables into the demand model as they would desire.

Finally, we recommend that the RCT be implemented at the market level $m$ if possible, rather than the individual level $i$. The reason is that a market-level RCT allows for maintaining market-level observables $X_m$ and $Z_m$ from equations \ref{eq: utility RCNL} and \ref{eq: gamma_i mu_i empirical}. However, running the experiments at the $m$ level is not always feasible, especially if the number of markets is small, or if the company is averse to exposing entire markets to treatments. In cases where market-level experiments are not available, individual-level RCTs should also work. In fact, due to feasibility constraints, our own application of the framework is presented on an individual-level RCT. However, one needs to remain cognizant of the limitations that come with running the RCT at the $i$ level: now different treatment groups will all share the same distribution over market characteristics. This means including characteristics $Z_m$ and $X_m$ is not feasible unless we change our market definition and define each market as the set of individuals who (i) are in the same treatment group, and (ii) share the same characteristics that we are interested in incorporating into $X_m,Z_m$. This solution, which is what we used in section \ref{subsec: more general empirical specification}, comes with caveats. Most importantly, our $X_m,Z_m$ measures have to be discrete (e.g., gender of the passenger, flight length below or above median, etc).

\section{Conclusion}

This paper developed a model of second-degree price discrimination (2PD). Our model follows the recent theoretical literature on 2PD, thereby allowing to replicate and build on some of the relevant results from that literature on how the joint distribution of the model primitives impacts the optimal selling mechanism. Among other objects, it captures the role of the covariance, across consumers, between the baseline valuation (i.e., affecting WTP for all versions of the product) and the differentiation valuation. Our results show that this covariance is key in determining the optimal 2PD mechanism. We argue, however, that this parameter is difficult to credibly estimate when the number of product versions is small (an typical characteristic in 2PD application), and when data on consumer characteristics that drive the covariance (e.g., possibly income) is not available to the mechanism designer. We then offer empirical tests that help estimate the parameters of the model under such constraints. Our partial-market tests fix the price of one version of the product, varying the other price. Our total-market tests vary both prices uniformly, shutting off cross substitution between versions. 

A key feature of our model is that it is not only tightly connected to the 2PD results from the mechanism design literature, but also connected to random-coefficients demand models (BLP). This means our framework allows to bring 2PD theoretical insights into empirical analysis of real 2PD applications. To demonstrate this, we implement the experiment design and estimate the empirical model in the context of seat selection, in collaboration with an international airline based in Asia. The differentiating feature between the two versions of the product in our context is timing of purchase. We estimate the model and find that differentiating the prices based on the timing of purchase has only a negligible impact on firm profit and consumer/social welfare. We also use this analysis to showcase the direct interpretability of our experiment design. We argue that the results from our partial- and total-market tests in this application are already suggestive that price discrimination would have little effect, even before a structural model of demand is estimated and the optimal pricing policy analyzed.

We further expand on our interpretability analysis by studying a number of simulated datasets presented in the form of market shares for the two products under the control group, partial-market test and total-market test. Each simulated dataset, hence, represents a scenario on how the market responds to the key tests. We argue that the propositions we developed when analyzing the optimal mechanisms, alongside our propositions for experiment design, can help interpret the results and form expectations on the shape of the optimal policy. We then estimate the model and solve for the optimal policy, demonstrating that those expectations are confirmed.

Although the differentiating feature between the product versions in our application was purchase timing, the methodology applies more broadly. The framework applies in the case of any form of differentiating factor between the products. We expect digital and technology products (e.g., cloud computing, Gen AI products, digital devices such as smartphones and laptops, etc) to constitute a large set of possible applications. The only requirement is that the two types of tests be feasible to implement.

There are a number of future directions in which our study may be extended. While some, such as extending the analysis to more than two product versions, are straightforward, others, such as extending the framework to oligopoly markets, will be more involved.

\bibliographystyle{chicago}
\bibliography{Main_2025_09_arXiv}

\newpage
\section*{Appendices}
\appendix

\section{Table of notations}

\begin{table}[ht]
\centering
\begin{tabular}{lll}
\toprule
\textbf{Notation} & \textbf{Interpretation} & \textbf{Notes / Normalization} \\
\midrule
$a$ & Utility intercept (choke price) & — \\
$b$ & Price sensitivity parameter & — \\
$\mu_i$ & Consumer $i$’s baseline valuation random effect & $\mu_i \sim N(0,\sigma_\mu^2)$ \\
$\sigma_\mu$ & Standard deviation of $\mu_i$ (heterogeneity) & — \\
$\gamma^0$ & Mean incremental value of premium version & 0 refers to the premium version \\
$\gamma^1$ & Covariance between $\mu_i$ and premium valuation effect & sign($\gamma_1$) drives price ranking \\
$p_k$ & Price of version $k \in \{1,2\}$ & — \\
$c_k$ & Marginal cost of version $k$ & — \\
$\varepsilon_{i,k}$ & IID Type I extreme‐value preference shock & Var$(\varepsilon_{i,k})=\pi^2/6$, normalized to 1 \\
\bottomrule
\end{tabular}
\caption{Notation used in the model and empirical design.}
\label{tab:notation}
\end{table}

\section{Proofs}

\subsection{Proof of \cref{prop: toy model}}
\label{prop: toy model (proof)}

We invoke the revelation principle and examine the space of direct mechanisms. In a direct mechanism, the monopolist commits to a selling mechanism which consists of a payment rule $t:\mathbb{R}\rightarrow\mathbb{R}$ and an allocation rule $x:\mathbb{R}\rightarrow\{1,2,\emptyset\}$. In words, the mechanism $(t,x)$ can take a consumer type $\mu_i$ and, as a function of that, determine (i) which of the three options (version $k=1$, version $k=2$, or the outside option) she receives, and (ii) how much she should pay the monopolist. This mechanism is announced to all consumers. Then each consumer $i$ ``announces type $\Tilde{\mu}_i$'' which, in principle, may or may not coincide with her true type $\mu_i$. A mechanism satisfies incentive compatibility (IC) if each consumer is weakly better off announcing her true type:

$$\forall \Tilde{\mu}_i,\mu_i: u_{i,x(\mu_i)}-t(\mu_i)\geq u_{i,x(\Tilde{\mu}_i)}-t(\Tilde{\mu}_i)$$

A mechanism satisfies individual rationality (IR) if the payoff to being truthful is weakly  larger than the outside option payoff (in our case, zero):

$$\forall \mu_i: u_{i,x(\mu_i)}-t(\mu_i)\geq0$$

Applying the revelation principle to our context implies that the optimal posted pricing mechanism is equivalent to the optimal direct mechanism that satisfies IC and IR. Next, we turn to proving the statements of the theorem.

We first show that under $\gamma^1<0$, optimal pricing involves no price discrimination. In other words, the direct mechanism allocates to each individual either version $k=2$ at a price $p^*_2$ or the outside option. Note that the posted-pricing version of this direct mechanism would be implemented by posting prices $p^*_2$ and  $p^*_{1}\geq p^*_2$: under such prices, and due to the vertical differentiation between the two versions, no customer would purchase $k=1$.

 To see this, why the direct mechanism would not allocate $k=1$ to anyone, assume the contrary: a non-zero-measure set of customers are assigned version $k=1$ at price $p^*_1$. There are two possibilities:

\begin{itemize}
    \item $p^*_2\leq p^*_{1}$: This means that there are $\mui$ and $\Tilde{\mui}$ such that: $x(\Tilde{\mui})=2$, $x(\mui)=1$, we have $p_{2}=t(\Tilde{\mui})<t(\mui)=p_{1}$. Such a mechanism violates the IC condition since $\mui$ would prefer the allocation for $\Tilde{\mui}$, which entails both a better product and a smaller payment. Thus, $p^*_2\leq p^*_{1}$ cannot hold.

    \item $p^*_2>p^*_{1}$: Consider two consumer types $\mui$ and $\Tilde{\mui}$ and assume $x(\Tilde{\mui})=2$, $x(\mui)=1$, $t(\Tilde{\mui})=p^*_{2}$, and $t(\mui)=p^*_{1}$. That is, consumer $\Tilde{\mui}$ is allocated $k=2$ at price $p^*_2$ while $\mui$ is allocated $k=1$ at price $p^*_{1}<p_2$. Applying IC twice, we get:

    $$u_{i,x(\mui)}-t(\mui)\geq u_{i,x(\Tilde{\mui})}-t(\Tilde{\mui})$$

    and 

    $$u_{\Tilde{i},x(\mui)}-t(\mui)\leq u_{\Tilde{i},x(\Tilde{\mui})}-t(\Tilde{\mui})$$

where $\Tilde{i}$ indexes consumer $\Tilde{\mui}$. Replacing from \cref{eq: toy model segment 1 utility}, \cref{eq: gamma_i mu_i}, and the allocation into the above inequalities, we get:

$$a+\mui-p^*_{1}\geq a+\mui(1+\gamma^1)+\gamma^0-p^*_2$$

and

$$a+\Tilde{\mui}(1+\gamma^1)+\gamma^0-p^*_{2}\geq a+\Tilde{\mui}-p^*_{1}$$

Simplifying, we obtain:

$$p^*_2-p^*_{1}\geq \mui\gamma^1+\gamma^0$$
 and
$$\Tilde{\mui}\gamma^1+\gamma^0\geq p^*_2-p^*_{1}$$

Adding the left sides together and further simplifying, we have:

$$\Tilde{\mui}\gamma^1\geq \mui\gamma^1$$

By $\gamma^1<0$, this means $\Tilde{\mui}\leq\mui$. In words, the type allocated version $k=1$ has to be the ``higher'' type. Next, note that by IR, we have:

$$a+\Tilde{\mui}(1+\gamma^1)+\gamma^0-p^*_{2}\geq 0$$

This, combined with $\Tilde{\mui}\leq\mui$, yields:

$$a+{\mui}(1+\gamma^1)+\gamma^0-p^*_{2}\geq 0$$

This latter inequality means that if the monopolist increased the price of $k=1$ to a number at or above $p^*_2$, all types $\mui$ who before the price increase purchased $k=1$, would now switch to purchasing $k=2$ at $p^*_2>p^*_{1}$ (i.e., none of those customers would be lost to the outside option). This would raise the firm's profit strictly, contradicting the optimality of the original mechanism. As a result, under $\gamma^1<0$, price-discriminating with $p^*_2>p^*_{1}$, although possible to implement with IC and IR, would not be optimal from the perspective of the monopolist. 

\end{itemize}

The above discussion completes the proof in the case of $\gamma^1<0$. We now turn to the case of $\gamma^1\geq 0$. To complete the proof in this case, we closely follow the analysis in \cite{ghili2023characterization}. Note that under $\gamma^1\geq0$, the setting without error terms, the type space of consumers exhibits a notion of monotonicity. More specifically, for any $\mui\geq \Tilde{\mui}$, we have:

$$a+\mui\geq a+\Tilde{\mui}$$

Also, by $\gamma^1\geq 0$, we get:

$$\gamma^0+\mui\gamma^1\geq \gamma^0+\Tilde{\mui}\gamma^1$$

In words, consumer $\mui$ has both a higher willingness to pay for $k=1$ than does $\Tilde{\mui}$ and has a higher willingness to pay for an upgrade from $k=1$ to $k=2$ than does $\Tilde{\mui}$. In addition to monotonicity, the demand system also satisfies quasi-concavity: given that the utilities are distributed uniformly, the profit function is quasi-concave in prices.

By the monotonicity and quasi-concavity properties mentioned above, the main result from \cite{ghili2023characterization} applies. According to this result, price discrimination is optimal if and only if the demand for version $k=1$, if sold alone (i.e., no option to buy $k=2$) and priced optimally, is strictly larger than the demand for version $k=2$, if sold alone and priced optimally.

We skip the derivations, but it is straightforward to show that if version $k=2$ is sold on its own, the optimal price will be given by:

$$p^{**}_{2}=\max\big(\frac{\sigmamu(1+\gamma^1)+\gamma^0+a}{2b},\frac{a+\gamma^0-\sigmamu(1+\gamma^1)}{b}\big)$$

and that the corresponding demand volume is:

$$D^{**}_2=\min\big(\frac{\sigmamu(1+\gamma^1)+\gamma^0+a}{4\sigmamu(1+\gamma^1)},1\big)$$

Likewise, it can be shown that the optimal sold-alone price and demand volume for $k=1$ are, respectively, given by:

$$p^{**}_{1}=\max\big(\frac{\sigmamu+a}{2b},\frac{a-\sigmamu}{b}\big)$$

and:

$$D^{**}_{1}=\min\big(\frac{\sigmamu+a}{4\sigmamu},1\big)$$

Ghili's condition for the optimality of price discrimination, hence, translates to $D^{**}_2<D^{**}_{1}$. Replacing into this inequality from the above and simplifying, we get:

\begin{equation}\label{eq: optimality of PD}
    D^{**}_2<D^{**}_{1}\Leftrightarrow 3\sigmamu > a >\frac{\gamma^0}{\gamma^1}
\end{equation}

Next, we turn to what the optimal prices $(p^*_{1},p^*_2)$ are when price discrimination is optimal for the monopolist. \cite{ghili2023characterization} shows that the optimal price $p^*_{1}$ in this case will be the same as the optimal sold-alone price $p^{**}_{1}$. He also shows that $p^*=p^{**}_{1}+\Delta p^*$ where $\Delta p^*$ is the optimal price that the monopolist would charge under a hypothetical where (i) all consumers are endowed with $k=1$ and (ii) the monopolist sells ``an upgrade to $k=2$'' and prices it optimally. Note that each consumer $i$'s utility for such an upgrade will be given by:

\begin{equation}\label{eq: diff utility}
    u_{i,diff}=u_{i,2}-u_{i,1}\equiv \gamma_i-b\Delta p
\end{equation}

where $\gamma_i:=\gamma^0+\gamma^1\mu_i$ is uniformly distributed on the interval $[\gamma^0-\gamma^1\sigmamu,\gamma^0+\gamma^1\sigmamu]$. This uniform distribution yields a linear demand curve, thereby allowing for a straightforward derivation of optimal pricing. We skip the derivation. The result is as follows:

$$\Delta p^*=\frac{\sigmamu\gamma^1+\gamma^0}{2b}$$

Note that both $p^*_{1}$ and $\Delta p^*$ agree with the claims made in the statement of the proposition.

Also note that for the case where non-discrimination is optimal, the optimal price $p^*_2$ is the same as $p^{**}$ derived above, which agrees with the statement of the proposition. This finishes the proof of the proposition. \textbf{Q.E.D.}

\subsection{Proof of \cref{prop: price discrimination error terms}}
\label{prop: price discrimination error terms (proof)}
We only show that $p^*_{1}>p^*_2$ if $\gamma^1<0$. The proofs of the other cases (i.e., $\gamma^1=0$ and $\gamma^1>0$) are identical. Also, for this proof, we will normalize the total market size to 1. This is without loss.

We take a contrapositive approach. We assume that $\gamma^1<0$ and, at the same time, $p_{1}^*\leq p^*_2$; and we reach a contradiction.

Start by observing that the share of product $k$ under arbitrary prices $(p_{1},p_2)$, conditional on the set of consumers who have the same level of \mui, is given by:

\begin{equation}\label{eq: proof share k}
    s_{k|\mui}(p_{1},p_2)=\frac{e^{a-bp_k+(\gamma^0+\mui\gamma^1)\mathds{1}_{k=2}+\mui}}{1+\Sigma_{k'}e^{a-bp_{k'}+(\gamma^0+\mui\gamma^1)\mathds{1}_{k'=2}+\mui}}
\end{equation}

The profit to the monopolist, $\pi(p_{1},p_2)$ is, hence, given by:

\begin{equation}\label{eq: profit based on shares}
    \pi(p_{1},p_2)=\Sigma_{k}\big(p_k\times\int_{\mui}s_{k|\mui}(p_{1},p_2) f_{\mui} d\mui\big)
\end{equation}

\begin{lemma}\label{lem: error terms prop, partial derivative formulas}
    At any pair of prices $(p_{1},p_2)$, the partial derivative of the profit function with respect to price $p_k$ of version $k$ is given by:

\begin{equation}\label{eq: lemma partial derivate of profit}
    \forall k\in\{1,2\}:\,\frac{\partial \pi}{\partial p_k}=  \int_{\mui}s_{k|\mui}\times\big(1-b\times p_k\times(1-s_{k|\mui})+b\times p_{k'}\times s_{k'|\mui}\big) f_{\mui} d\mui
\end{equation}

where $k':=3-k$ is the other version, and where the notation on the dependency of profits and shares has been suppressed for brevity.

Also, the sum of the derivatives of the profit with respect to all prices is given by:

\begin{equation}\label{eq: lemma total derivate of profit}
    \Sigma_{k}\frac{\partial \pi}{\partial p_k}=\Sigma_k  \big(\int_{\mui}s_{k|\mui} f_{\mui} d\mui\big) -b\Sigma_k \big( p_k\times \int_{\mui}s_{k|\mui}\times s_{\emptyset|\mui} f_{\mui}d\mu\big)
\end{equation}

where $s_{\emptyset}$ denotes the share of the outside option.

\end{lemma}

\textbf{Proof.} The proof of this lemma is tedious but straightforward and involves replacing from \cref{eq: proof share k} into \cref{eq: profit based on shares}, differentiating, and re-arranging terms. It is left to the reader.

Now, observe that by the first order condition at the optimal prices $(p^*_{1},p^*_2)$, the gradient of the profit function with respect to the vector of prices is zero. That is:

\begin{equation}\label{eq: total derivative zero at optimal}
    \Sigma_{k}\frac{\partial \pi}{\partial p_k}|_{(p_{1}^*,p^*_2)}=0
\end{equation}

We will use \cref{eq: total derivative zero at optimal} to show that $\frac{\partial \pi}{\partial p_{1}}|_{(p_{1}^*,p^*_2)}>0$, which would contradict the optimality of $(p^*_{1},p^*_2)$. Before we can do show this, we need additional lemmas.

\begin{lemma}\label{lem: MLR}
    Take any pair $\mui>\mui'$. Under  any pair of price $(p_{1},p_2)$, optimal or otherwise, the following are true when $\gamma^1<0$:

    \begin{itemize}
        \item If we denote by $s_{\mui}:=\Sigma_k s_{k|\mui}$, we have: $$\frac{s_{1|\mui}}{s_{1|\mui'}}>\frac{s_{\mui}}{s_{\mui'}}>\frac{s_{2|\mui}}{s_{2|\mui'}}$$

        \item If $\gamma^1>-1$, then $s_{\mui}>s_{\mui'}$.
        
    \end{itemize}
\end{lemma}

\begin{lemma}\label{lem: error terms prop: intermediate inequality}
    At the optimal prices, we have: 
    $$\int_{\mui}s_{1|\mui}\times (1-bp^*_{1}\times s_{\emptyset|\mui})f_{\mui}d\mui>0.$$
\end{lemma}

We first use these lemmas to complete the proof of the main proposition by reaching the contradiction. We will then provide the proofs of these lemmas (note that the proofs of the lemmas may use the contrapositive assumption of $p^*_2\geq p^*_1$ ).

To reach a contradiction, we will show that at the optimal prices, we get $\frac{\partial \pi}{\partial p_{1}}>0$, which contradicts the optimality. To see why $\frac{\partial \pi}{\partial p_{1}}>0$ holds, observe that \cref{eq: lemma partial derivate of profit} can be rearranged as:

    $$\frac{\partial \pi}{\partial p_k}=  \int_{\mui}s_{k|\mui}\times\big(1-b\times p_k\times(1-s_{k|\mui}-s_{k'|\mui})\big) f_{\mui} d\mui$$

\begin{equation}\label{eq: lemma partial derivate of profit rearranged 1}
    + \int_{\mui}s_{k|\mui}\times b\times\big( p_{k'}-p_k\big)\times s_{k'|\mui} f_{\mui} d\mui
\end{equation}

Note that in this rearrangement, $\frac{\partial \pi}{\partial p_k}$ has two additive terms. Once we set $k=1$ and observe that $1-s_{k|\mui}-s_{k'|\mui}$ is by definition the same object as $s_{\emptyset|\mui}$ the first term in the expression for $\frac{\partial \pi}{\partial p_{1}}$ collapses to the term that \cref{lem: error terms prop: intermediate inequality} proves to be strictly positive. The second term is non-negative since it is an integral over a multiplication of four objects each of which is non-negative: shares $s_{1|\mui}$ and $s_{2|\mui}$ are by construction non-negative. So is the price coefficient $b$. As for $p_{k'}-p_k$, note that we are evaluating this term under the optimal prices and under $k=1$. This term, hence, boils down to $p^*_2-p^*_{1}$, which by our contrapositive assumption, is non-negative. As a result, $\frac{\partial \pi}{\partial p_{1}}$ should be strictly positive when evaluated at optimal prices. This contradicts the optimality assumption, thereby proving the proposition assuming lemmas \ref{lem: MLR} and \ref{lem: error terms prop: intermediate inequality} are correct. We turn to proving these two lemmas next.

\textbf{Proof of \cref{lem: MLR}.} It is straightforward to derive the following expression for $\frac{s_{2,\mui}}{s_{1,\mui}}$:

$$\frac{s_{2,\mui}}{s_{1,\mui}}=e^{\gamma-b\times (p_2-p_{1})}$$

where $\gamma:=\gamma^0+\gamma^1\times \mu$.

Thus, for any pair $\mui$ and $\mui'$, we get:

$$\frac{s_{2,\mui}}{s_{1,\mui}}\mathrel{\big/}\frac{s_{2,\mui'}}{s_{1,\mui'}}=e^{\gamma-\gamma'}=e^{(\mui-\mui')\times \gamma^1}$$

As a result, if $\gamma^1<0$ and $\mui>\mui'$, we get:

$$\frac{s_{2,\mui}}{s_{1,\mui}}\mathrel{\big/}\frac{s_{2,\mui'}}{s_{1,\mui'}}<1$$

Re-arranging, we get:

$$\frac{s_{2,\mui}}{s_{2,\mui'}}<\frac{s_{1,\mui}}{s_{1,\mui'}}$$

which is part of the first statement of the lemma. Noticing that $s_{\mui}=s_{1|\mui}+s_{2|\mui}$ and $s_{\mui'}=s_{1|\mui'}+s_{2|\mui'}$, it is straightforward (hence left to the reader) to verify that the three-way comparison also holds:

$$\frac{s_{2,\mui}}{s_{2,\mui'}}<\frac{s_{\mui}}{s_{\mui'}}<\frac{s_{1,\mui}}{s_{1,\mui'}}$$

We now turn to the second statement of the lemma. With $\gamma^1\geq -1$, one can verify that for any $\mui>\mui'$, we have:

$$a-bp_{1}+\mui>a-bp_{1}+\mui'$$

and 

$$a-bp_{2}+\mui+\gamma^0+\gamma^1\times\mui\geq a-bp_{2}+\mui'+\gamma^1\times\mui'$$

In other words: $\mathbb{E}u_{k|\mui}\geq\mathbb{E}u_{k|\mui'}$ for all $k$, and the inequality is strict for $k=1$. As a result:

$$\Sigma_k e^{\mathbb{E}u_{k|\mui}}>\Sigma_k e^{\mathbb{E}u_{k|\mui'}}$$

$$\Rightarrow \frac{\Sigma_k e^{\mathbb{E}u_{k|\mui}}}{1+\Sigma_k e^{\mathbb{E}u_{k|\mui}}}
>
\frac{\Sigma_k e^{\mathbb{E}u_{k|\mui'}}}{1+\Sigma_k e^{\mathbb{E}u_{k|\mui'}}}$$

which means $s_{\mui}>s_{\mui'}$.

Both claims in the lemma have now been proven. $\blacksquare$

\textbf{Proof of \cref{lem: error terms prop: intermediate inequality}.} Start by recalling that the optimality of $(p_{1}^*,p^*_2)$ implies $\Sigma_{k}\frac{\partial \pi}{\partial p_k}=0$. Replacing from \cref{eq: lemma total derivate of profit}, we obtain:

$$\Sigma_k  \big(\int_{\mui}s_{k|\mui} f_{\mui} d\mui\big) -b\Sigma_k \big( p^*_k\times \int_{\mui}s_{k|\mui}\times s_{\emptyset|\mui} f_{\mui}\big)=0$$

Replacing $p^*_2$ by $p^*_{1}$ in the above formula and recalling the contrapositive assumption $p^*_2\geq p^*_{1}$, we get:

$$\Sigma_k  \big(\int_{\mui}s_{k|\mui} f_{\mui} d\mui\big) -b\Sigma_k \big( p^*_{1}\times \int_{\mui}s_{k|\mui}\times s_{\emptyset|\mui} f_{\mui}\big)\geq 0$$

Further rearranging, we get:

$$\Sigma_k  \big(\int_{\mui}s_{k|\mui} \times(1-bp^*_{1}\times s_{\emptyset|\mui}) f_{\mui} d\mui\big)\geq 0$$

Denoting  $s_{\mui}:=\Sigma_k s_{k|\mui}$, we get:

\begin{equation}\label{eq: lem  error terms prop interim inequality}
    \int_{\mui}s_{\mui}\times (1-bp^*_{1}\times s_{\emptyset|\mui}) f_{\mui} d\mui\geq 0
\end{equation}

Note that \cref{eq: lem  error terms prop interim inequality} bears similarities to the statement of the lemma which we seek to prove. There are two differences, however: (i) the inequality in the lemma is strict; (ii) the inequality in the lemma has an $s_{1|\mu}$ term whereas  \cref{eq: lem  error terms prop interim inequality} contains an $s_{\mu}$ term instead. Thus, it takes additional operations to prove the lemma from \cref{eq: lem  error terms prop interim inequality}, which we carry out below.

Denote by $\Tilde{\mu}$ a value of $\mui$ such that $1-bp^*_{1}\times s_{\emptyset|\Tilde{\mu}}=0$. If such $\Tilde{\mu}$ cannot be chosen because  $1-bp^*_{1}\times s_{\emptyset|{\mui}}$ is strictly positive for the entire support of $\mui$, then set $\Tilde{\mu}$ to be the largest number in the support. Then, for any $\mui > (<) \Tilde{\mu}$, we have:

\begin{equation}\label{eq: lem  error terms prop auxilary inequality}
    1-bp^*_{1}\times s_{\emptyset|{\mui}} > (<) 0
\end{equation}

and

\begin{equation}\label{eq: lem  error terms prop auxilary inequality 2}
    \frac{s_{1|\mui}}{s_{\mui}}\mathrel{\big/}\frac{s_{1|\Tilde{\mu}}}{s_{\Tilde{\mu}}}>(<)1
\end{equation}

where the latter inequality directly follows from \cref{lem: MLR}. Note that \cref{eq: lem  error terms prop auxilary inequality} and \cref{eq: lem  error terms prop auxilary inequality 2}, together, imply that for any $\mui$ we have:

\begin{equation}\label{eq: lem  error terms prop auxilary inequality 3}
    1-bp^*_{1}\times s_{\emptyset|{\mui}} > (<) 0\Rightarrow    \frac{s_{1|\mui}}{s_{\mui}}\mathrel{\big/}\frac{s_{1|\Tilde{\mu}}}{s_{\Tilde{\mu}}}>(<)1
\end{equation}

Next, note that if we take \cref{eq: lem  error terms prop interim inequality}, multiply all the positive terms in the integral by a scalar strictly larger than 1, and multiply all the negative terms in the integral by a scalar strictly smaller than 1, the resulting expression will strictly increase, which means it will be strictly positive. Based on \cref{eq: lem  error terms prop auxilary inequality 3}, one can do this by multiplying the integrand by $    \frac{s_{1|\mui}}{s_{\mui}}\mathrel{\big/}\frac{s_{1|\Tilde{\mu}}}{s_{\Tilde{\mu}}}$. This yields:

$$\int_{\mui}(\frac{s_{1|\mui}}{s_{\mui}}\mathrel{\big/}\frac{s_{1|\Tilde{\mu}}}{s_{\Tilde{\mu}}})\times s_{\mui} \times(1-bp^*_{1}\times s_{\emptyset|\mui}) f_{\mui} d\mui> 0$$

Canceling the $s_{\mui}$ terms and factoring the constant term $\frac{s_{1|\Tilde{\mu}}}{s_{\Tilde{\mu}}}$ outside of the integral, we get:

$$\frac{s_{\Tilde{\mu}}}{s_{1|\Tilde{\mu}}}\times 
\int_{\mui}s_{1|\mui}\times(1-bp^*_{1}\times s_{\emptyset|\mui}) f_{\mui} d\mui> 0$$

By $\frac{s_{\Tilde{\mu}}}{s_{1|\Tilde{\mu}}}\geq 0$, we get:

$$
\int_{\mui}s_{1|\mui}\times(1-bp^*_{1}\times s_{\emptyset|\mui}) f_{\mui} d\mui> 0$$

which completes the proof of the lemma. $\blacksquare$

With the proofs of these two lemmas at hand, the proof of the proposition is complete. \textbf{Q.E.D.}

\subsection{Proofs of   \cref{prop: differential response rates}}
\label{prop: total market test (proof)}

Similar to \cref{prop: price discrimination error terms}, we prove this result for the case of $\gamma^1<0$. The $\gamma^1>0$ and $\gamma^1=0$ cases would be almost identical. In the rest of the proof, we only work with price vectors in the form of $(p_{1}+r,p_{2}+r)$ where $(p_{1},p_{2})$ is fixed and $r$ varies. As a result, we suppress the notations on the former and write various objects as functions of the latter only.

\begin{lemma}\label{lem: identification prop, eta}
    As we vary $r$, the relative shares of the two products $k=1$ and $k=2$ remain unchanged for any fixed level of $\mu$. Formally, there is a function $\eta:\mathbb{R}\rightarrow\mathbb{R}$ such that  for any $r\in\mathbb{R}:$ and any $\mui$, we have:

    $$\frac{s_{1|\mui}(r)}{s_{\mui}(r)}=\eta(\mui)$$

    This function is strictly increasing in $\mui$ when $\gamma^1<0$, constant when $\gamma^1=0$, and strictly decreasing when $\gamma^1>0$.
\end{lemma}

In other words, the $k=1$ to $k=2$ share ratio does not depend on the price shifter $r$.

\textbf{Proof.} Writing out the expressions for the shares, one can show:

$$\frac{s_{1|\mui}(r)}{s_{\mui}(r)}=\frac{1}{1+e^{\big(b(p_{1}-p_2)+\gamma^0+\gamma^1\times \mui\big)}}$$

It immediately follows that the expression is not a function of $r$ and depends only on $\mu$ and other fixed primitives. We can call this function $\eta(\mu)$. We also leave the verification of the monotonicity of $\eta(\mu)$, and the direction of this monotonicity as a function of the sign of $\gamma^1$, to the reader. $\blacksquare$

\begin{lemma}\label{lem: MLR inequality for the identification prop}
    Assume the function $\eta(\mui)$ is positive and strictly increasing in $\mui$. Also take two probability distributions $w_1(\cdot)$ and $w_2(\cdot)$  over $\mui$ that have the same support $[\underline{\mu},\bar{\mu}]$ and assume they are both strictly positive on this support. If $w_1$ dominates $w_2$ in a Monotone Likelihood Ratio (MLR) sense: $\forall \mui>\mui':\,\frac{w_1(\mui)}{w_1(\mui')}>\frac{w_2(\mui)}{w_2(\mui')}$, then we have:

    $$\int_{\mui}\eta(\mui)w_1(\mui)d\mui>\int_{\mui}\eta(\mui)w_2(\mui)d\mui$$
\end{lemma}

\textbf{Proof.} Given that $w_1$ and $w_2$ are probability distributions, we get: $\int_{\mui}w_1(\mui)d\mui=\int_{\mui}w_2(\mui)d\mui$. This, combined with the MLR relationship, implies that:

$$\exists\Tilde{\mu}:\begin{cases}
  w_1(\mui)>w_2(\mui) & \forall \mui>\Tilde{\mu} \\
  w_1(\mui)<w_2(\mui) & \forall \mui<\Tilde{\mu}
\end{cases}$$

Denoting $\Delta w(\mui):=w_1(\mui)-w_2(\mui)$, we can write: $\int_{\mui}\Delta w(\mui) d\mui=0$. Multiplying by $\eta(\Tilde{\mu})$, we obtain:

$$\int_{\mui}\eta(\Tilde{\mu})\Delta w(\mui) d\mui=0$$

$$\Rightarrow\int^{\tilde{\mu}}_{\mui=\underline{\mu}}\eta(\Tilde{\mu})\Delta w(\mui) d\mui+
\int^{\bar{\mu}}_{\mui=\tilde{\mu}}\eta(\Tilde{\mu})\Delta w(\mui) d\mui=0$$

Note that the first integral consists of strictly negative $\Delta w$ terms while the second integral consists of strictly positive ones. As a result, if we discount the terms in the first integral by multiplying them by factors in $(0,1)$ and amplify the terms in the second by multiplying its terms by factors in $(1,\infty)$, the expression will strictly increase. Next, observe that by $\eta$ being positive and strictly increasing, the fraction $\frac{\eta(\mui)}{\eta(\tilde{\mu})}$ is in $(0,1)$ for all $\mui<\tilde{\mu}$ and it is in $(1,\infty)$ for all $\mui>\tilde{\mu}$. Thus, we can write:

$$\int^{\tilde{\mu}}_{\mui=\underline{\mu}}\eta(\Tilde{\mu})\frac{\eta(\mui)}{\eta(\tilde{\mu})}\times\Delta w(\mui) d\mui+
\int^{\bar{\mu}}_{\mui=\tilde{\mu}}\frac{\eta(\mui)}{\eta(\tilde{\mu})}\times\eta(\Tilde{\mu})\Delta w(\mui) d\mui>0$$

Simplifying and re-arranging, we get:

$$\int^{\tilde{\mu}}_{\mui=\underline{\mu}}\eta(\mui)\times\Delta w(\mui) d\mui+
\int^{\bar{\mu}}_{\mui=\tilde{\mu}}\eta(\mui)\times\eta(\Tilde{\mu})\Delta w(\mui) d\mui>0$$

$$\Rightarrow\int^{\bar{\mu}}_{\mui=\underline{\mu}}\eta(\mui)\times\Delta w(\mui) d\mui>0$$

$$\Rightarrow\int^{\bar{\mu}}_{\mui=\underline{\mu}}\eta(\mui)\times w_1(\mui) d\mui>\int^{\bar{\mu}}_{\mui=\underline{\mu}}\eta(\mui)\times w_2(\mui) d\mui$$

which is the statement of the lemma.  $\blacksquare$

\begin{lemma}\label{lem: last lemma in the proof of prop for ratios}
    For any $r>0$, the following comparison holds relative to $r=0$:

    $$\frac{s_{1}(r)}{s(r)}>\frac{s_{1}(0)}{s(0)}$$

where the term $s$ denotes the total ($k=1$ and $k=2$) share.
\end{lemma}

\textbf{Proof.} For any $r$, the expression $\frac{s_{1}(r)}{s(r)}$ can be expanded as:

$$\frac{s_{1}(r)}{s(r)}=\frac{\int_{\mui}s_{1|\mui}(r)f(\mui)d\mui}{\int_{\mui}s_{\mui}(r)f(\mui)d\mui}$$

By lemma \ref{lem: identification prop, eta} and the definition of the function $\eta$, we get:

$$\frac{s_{1}(r)}{s(r)}=\frac{\int_{\mui}\eta(\mui)\times s_{\mui}(r)f(\mui)d\mui}{\int_{\mui}s_{\mui}(r)f(\mui)d\mui}$$

Bringing the denominator inside the integral in the numerator yields:

$$\frac{s_{1}(r)}{s(r)}=\int_{\mui}\eta(\mui)\times\frac{ s_{\mui}(r)f(\mui)}{\int_{\tilde{\mui}}s_{\tilde{\mui}}(r)f(\tilde{\mui})d\tilde{\mui}}d\mui$$

Next, denote $\frac{ s_{\mui}(r)f(\mui)}{\int_{\tilde{\mui}}s_{\tilde{\mui}}(r)f(\tilde{\mui})d\tilde{\mui}}$ by $w_r(\mui)$.

If we follow similar steps to the above for the $r=0$ case and define the function $w_0(\mui)$ appropriately, then proving the lemma would be equivalent to proving the following:

\begin{equation}\label{eq: last lemma in the ratios prop proof}
   \int_{\mui}\eta(\mui)w_r(\mui)d\mui>\int_{\mui}\eta(\mui)w_0(\mui)d\mui 
\end{equation}

If we show that $w_0$ and $w_r$ are both probability distributions and that the latter dominates the former in the MLR sense, we can apply lemma \ref{lem: MLR inequality for the identification prop} and proof the current lemma.

Verifying that these functions are probability distributions is straightforward. One could integrate over $\mui$ and show the outcome is 1:

$$\int_{\mui}w_r(\mui)d\mui=\int_{\mui}\frac{ s_{\mui}(r)f(\mui)}{\int_{\tilde{\mui}}s_{\tilde{\mui}}(r)f(\tilde{\mui})d\tilde{\mui}}d\mui$$

$$=\frac{\int_{\mui} s_{\mui}(r)f(\mui)d\mui}{\int_{\tilde{\mui}}s_{\tilde{\mui}}(r)f(\tilde{\mui})d\tilde{\mui}}=1$$

Clearly, the same applies when $r=0$.

We next show that for any $\mui>\mui'$:

$$\frac{w_r(\mui)}{w_r(\mui')}>\frac{w_0(\mui)}{w_0(\mui')}$$

It is straightforward to verify that all $f(\mui)$ and $f(\mui')$ terms, as well as the integrals in the denominators of these functions, cancel out, and the above comparison becomes equivalent to the following:

$$\frac{s_{\mui}(r)}{s_{\mui'}(r)}>\frac{s_{\mui}(0)}{s_{\mui'}(0)}$$

or, equivalently:

$$\frac{s_{\mui}(r)}{s_{\mui}(0)}>\frac{s_{\mui'}(r)}{s_{\mui'}(0)}$$

If for any $k$ we denote $v_k(\mui):=a-b\times p_k+\mui+\mathds{1}_{k=2}\times(\gamma^0+\gamma^1\mui)$, (i.e., basically  the expected utility for product $k$), one can show the following for any $\mui$:

$$\frac{s_{\mui}(r)}{s_{\mui}(0)}=\frac{1+\Sigma_{k}e^{v_k(\mui)}}{e^{br}+\Sigma_{k}e^{v_k(\mui)}}$$

Next, observe that by $\mui>\mui'$ and $|\gamma^1|<1$, we have: $\forall k: v_k(\mui)>v_k(\mui')$. This, in turn, implies:

$$\Sigma_ke^v_k(\mui)>\Sigma_ke^v_k(\mui')$$

This latter inequality, combined with $e^{br}>1$ (which comes from $b>0$ and $r>0$), yields:

$$\frac{1+\Sigma_{k}e^{v_k(\mui)}}{e^{br}+\Sigma_{k}e^{v_k(\mui)}}>\frac{1+\Sigma_{k}e^{v_k(\mui')}}{e^{br}+\Sigma_{k}e^{v_k(\mui')}}$$

Thus, we have $\frac{s_{\mui}(r)}{s_{\mui}(0)}>\frac{s_{\mui'}(r)}{s_{\mui'}(0)}$, which means probability distribution $w_r$ strictly MLR-dominates probability distribution $w_0$.  Thus, lemma \ref{lem: MLR inequality for the identification prop} applies to \cref{eq: last lemma in the ratios prop proof} and the proof of the current lemma is now complete. $\blacksquare$.

With lemma \ref{lem: last lemma in the proof of prop for ratios} proven, the proposition immediately follows from the observation that $\forall r: s(r):=s_{1}(r)+s_2(r)$:

 $$\frac{s_{1}(r)}{s(r)}>\frac{s_{1}(0)}{s(0)}$$

 $$\Rightarrow \frac{s_{1}(r)}{s_2(r)}>\frac{s_{1}(0)}{s_2(0)}$$

\textbf{Q.E.D.}

\section{Extension to marginal costs}\label{apx: marginal costs}

In this section, we state and prove a version of \cref{prop: price discrimination error terms} that allows for nonzero and heterogeneous marginal costs across versions.

\begin{proposition}\label{prop: marginal costs}
    Suppose the conditions for \cref{prop: price discrimination error terms} hold. Also, suppose the marginal costs for producing version $k$ are given by $c_k\geq 0$. The following are true about the monopolist's optimal prices $(p^*_{1},p^*_2)$:
    \begin{itemize}
        \item If $\gamma^1<0$: then $p^*_{1}-c_{1}>p^*_2-c_{2}$.
        \item If $\gamma^1>0$: then $p^*_{1}-c_{1}<p^*_2-c_{2}$.
    \end{itemize}
\end{proposition}

In words, \cref{prop: marginal costs} shows that the only change one needs to make to the statement of \cref{prop: price discrimination error terms} under non-zero marginal costs is to replace prices with margins.

\textbf{Proof.} Consider an ``auxiliary'' statement of a different 2PD problem with the following changes relative to what we have in the statement of the proposition:

\begin{enumerate}
    \item Marginal costs are zero
    \item Instead of \cref{eq: gamma_i mu_i,eq: theory utility error terms}, the utility is described by the following equations:    
\end{enumerate}

\begin{equation}\label{eq: theory utility error terms margin}
    u_{ik}=(a-b\times c_{1})-b\times p_k+\mu_i+\gamma_i\times\mathds{1}_{k=2}+\epsilon_{ik}
\end{equation}

and

\begin{equation}\label{eq: gamma_i mu_i margin}
    \gamma_i=\big(\gamma^0-b\times (c_{2}-c_{1})\big)+\gamma^1\times \mu_i
\end{equation}

In words, this new problem sets marginal costs back to zero but replaces the utility parameters $a$ and $\gamma^0$, respectively, with $a-b\times c_{1}$ and $\gamma^0-b\times (c_{2}-c_{1})$.

\begin{lemma}\label{lem: equivalence auxilary original}
If a monopolist facing the auxiliary problem charges prices $(m_{1},m_2)$, she will have the same demand (for each version) and the same total profit as a monopolist who faces the original problem and charges prices $(p_{1},p_2)=(m_{1}+c_{1},m_2+c_{2})$.
    
\end{lemma}

\textbf{Proof.} To see why the demands are the same, observe that all the utilities are the same. In one replaces $m_k=p_k-c_k$ into the formulas \cref{eq: gamma_i mu_i margin,eq: theory utility error terms margin} and expands, one can verify that they collapse to the original utility formulas \cref{eq: gamma_i mu_i,eq: theory utility error terms}. Next, observe that in addition to all demand levels, margins are also the same between the two problems. In both cases, the margins are $m_{1}$ and $m_2$. Thus, profits are the same as well. $\blacksquare$

Next, note that (i) \cref{prop: price discrimination error terms} is applicable to the auxiliary problem, and that (ii) the formulation of the auxiliary problem modified $a$ and $\gamma^0$ but did not change   
 $\gamma^1$. Applying \cref{prop: price discrimination error terms}, hence, yields:

    \begin{itemize}
        \item If $\gamma^1<0$: then $m^*_{1}>m^*_2$.
        \item If $\gamma^1>0$: then $m^*_{1}<m^*_2$.
    \end{itemize}

where $m_{1}^*$ and $m^*_2$ are, respectively, the optimal prices for versions $k=1$ and $k=2$ in the auxiliary problem. But given \cref{lem: equivalence auxilary original}, $m^*_{1}>m^*_2$ is equivalent to $p^*_{1}-c^*_{1}>p^*_{2}-c^*_{2}$, and the similar equivalences apply when $m^*_{1}<m^*_2$ or $m^*_{1}=m^*_2$. This completes the proof of \cref{prop: marginal costs}. \textbf{Q.E.D.}

Finally, it is worth noting that the equivalent of \cref{cor: data-->mechanism}, which connected data patterns directly to optimal policy under zero marginal costs, can also be stated in the case of marginal costs.

\begin{corollary}\label{cor: data-->mechanism; MC}

Suppose the conditions in \cref{prop: marginal costs,prop: differential response rates} hold. Then, we have:
    \[
\phi_{1,2}\, {\underset{(<)}{>}} \,1 \Rightarrow p^*_2-c_2\, {\underset{(<)}{>}}\, p^*_{1}-c_1
\]
\end{corollary}

In other words, even under marginal costs, one can leverage the outcome of a differential response treatment in order to obtain qualitative insights about optimal pricing policy.

\end{document}